\documentclass[usenatbib]{mnras}

\usepackage{newtxtext}%,newtxmath}
\usepackage[T1]{fontenc}
\usepackage{ae,aecompl}
\usepackage{placeins}
\usepackage{graphicx}	% Including figure files
\usepackage{amsmath}	% Advanced maths commands
\usepackage{amssymb}	% Extra maths symbols

\newcommand{\lcdm}{CDM }

\newcommand{\chii}{$\chi^2_{\rm init}$}
\newcommand{\chif}{$\chi^2_{\rm init} = 50\%$}
\newcommand{\chih}{$\chi^2_{\rm init} = 100\%$}
\newcommand{\kms}{km~s$^{-1}$}
\defcitealias{Vogelsberger19v484}{V19}
\defcitealias{Hansen2006}{HM06}

\title[Impact of inelastic SIDM on a MW halo]{The impact of inelastic self-interacting dark matter on the dark matter structure of a Milky Way halo}

\author[K.T.E. Chua et. al.]{
Kun Ting Eddie Chua$^{1}$\thanks{Email: eddiechua@ihpc.a-star.edu.sg},
Karia Dibert$^{2}$,
Mark Vogelsberger$^{2}$,
and Jes{\'u}s Zavala$^{3}$\\
$^{1}$Institute of High Performance Computing, 1 Fusionopolis Way \#16-01, Singapore 138632\\
$^{2}$Department of Physics, Massachusetts Institute of Technology, 77 Massachusetts Avenue, Cambridge, MA 02139, USA\\
$^{3}$Center for Astrophysics and Cosmology, Science Institute, University of Iceland, Dunhagi 5, 107 Reykjavik, Iceland
}

\date{Accepted XXX. Received YYY; in original form ZZZ}
\pubyear{2020}

\begin{document}

\label{firstpage}
\pagerange{\pageref{firstpage}--\pageref{lastpage}}
\maketitle

\begin{abstract}
We study the effects of inelastic dark matter self-interactions on the internal structure of a simulated Milky Way (MW)-size halo. 
Self-interacting dark matter (SIDM) is an alternative to collisionless cold dark matter (CDM) which offers a unique solution to the problems encountered with CDM on sub-galactic scales.
Although previous SIDM simulations have mainly considered elastic collisions, theoretical considerations motivate the existence of multi-state dark matter where transitions from the excited to the ground state are exothermic.
In this work, we consider a self-interacting, two-state dark matter model with inelastic collisions, implemented in the {\sc Arepo} code.
We find that energy injection from inelastic self-interactions reduces the central density of the MW halo in a shorter timescale relative to the elastic scale, resulting in a larger core size.
Inelastic collisions also isotropize the orbits, resulting in an overall lower velocity anisotropy for the inelastic MW halo.
In the inner halo, the inelastic SIDM case (minor-to-major axis ratio $s \equiv c/a \approx 0.65$) is more spherical than the CDM ($s \approx 0.4$), but less spherical than the elastic SIDM case ($s \approx 0.75$).
The speed distribution $f(v)$ of dark matter particles at the location of the Sun in the inelastic SIDM model shows a significant departure from the CDM model, with $f(v)$ falling more steeply at high speeds.
In addition, the velocity kicks imparted during inelastic collisions produce unbound high-speed particles with velocities up to 500~km~s$^{-1}$ throughout the halo.
This implies that inelastic SIDM can potentially leave distinct signatures in direct detection experiments, relative to elastic SIDM and CDM.
\end{abstract}

\begin{keywords}
cosmology: dark matter -- galaxies: haloes -- methods: numerical
\end{keywords}

%%%%%%%%%%%%%%%%% BODY OF PAPER %%%%%%%%%%%%%%%%%%
\section{Introduction}

The standard cold dark matter (CDM) model has been successful in describing several properties of the universe, such as the Cosmic Microwave Background, and the formation and evolution of large-scale structure in the distribution of galaxies throughout the universe \citep[e.g.][]{Spergel2003, Springel2005}. 
On the other hand, relevant discrepancies have been uncovered between CDM and observations at sub-galactic scales.
%\citep{Bullock2017}.
Some of these include:
(i) the cusp-core problem, arising from a disagreement between the predicted dark matter (DM) density profiles of low-mass galaxies and their observed density profiles \citep[e.g.][]{deBlok1997,Walker2011};
(ii) the missing satellites problem, arising from a discrepancy between the abundance of low-mass galaxies in the Milky Way (MW) and that ``naively" predicted by CDM simulations (e.g. \citealt{Moore1999}; a similar problem appears in field galaxies, e.g. \citealt{Zavala2009,Klypin15v454});
(iii) the too-big-to-fail problem highlighted by \cite{BK2011}, which arises from the fact that low-mass CDM (sub)haloes, which are expected to host low-mass galaxies, are seemingly too dense to explain the observed stellar kinematics of these systems;
(iv) the dwarf rotation curve diversity problem, where the rotation curves of simulated dwarf galaxies at a fixed mass do not show the large variety/diversity observed in real galaxies \citep{Oman2015}\footnote{
We note that with recent observations of ultra-faint galaxies, the too-big-to-fail problem also becomes a diversity problem for the broad distribution of stellar kinematics in dwarf spheroidals in the MW \citep{Zavala19v100}}; and
(v) conflicting expectations of the dark matter halo shape, where \lcdm simulations predict average shapes far less spherical than those derived from tidal stream observations \citep[e.g.][]{Ibata01v551,Law10v714,Vera-Ciro13v773,Bovy16v833}. 

It is important to remark that these small-scale disagreements are only firmly established from the results of $N$-body simulations, which include only the effects of DM gravity.
Recent efforts have accelerated the development of hydrodynamics simulations, e.g. Illustris \citep{Vogelsberger2014a}, EAGLE \citep{EAGLE}, Illustris-TNG \citep{Pillepich2018a}, where the incorporation of galaxy formation processes allows the complex modelling of galaxies to occur in a full cosmological context.
Due to the coupling between DM and baryons through gravity, it has been suggested that baryonic effects in hydrodynamic simulations can alleviate some, if not all, of the CDM challenges.
For example, energy from supernova feedback can inject energy into the inner dark matter halo, alleviating the cusp-core problem \citep{Navarro1996,Governato2012,Onorbe2015,Read2016}, as well as the too-big-to-fail problem \citep{Zolotov2012,Brook2015,Sawala2016,Wetzel2016}.
It has also been found that while a combination of supernova and active galactic nuclei (AGN) feedback can reduce the central densities of haloes \citep[e.g.][]{Duffy10v405}, the condensation of baryons in halo centres conversely increases the halo concentration through adiabatic contraction \citep{Blumenthal86v301,Gnedin04v616}.
The condensation of baryons can also affect halo shapes, leading to more spherical DM haloes in  hydrodynamic simulations compared to $N$-body simulations \citep[e.g.][]{Katz91v377,Dubinski94v431,Tissera10v406,Abadi10v407,Bryan13v429,Butsky16v462,Chisari2017,Chua19v484}. 
The impact of all of these baryonic processes certainly reduces the tension between the CDM model and observations of the dwarf galaxy population. 
However, it is important to note that a wide variety of physics implementations exist, resulting in uncertain predictions.
If these uncertainties in baryonic physics must be taken into account, we argue that is also prudent to consider another uncertainty in the physics of galaxies, which is the nature of the dark matter particle. 
Alternatives to dark matter models can arise by relaxing the key assumptions of CDM - that dark matter is both cold and collisionless.
For a recent review on different dark matter models and their impact on structure formation, see \cite{Zavala19v7}; for a review of  the  CDM  challenges  and  possible solutions, see \cite{Bullock17v55}.

Relaxing the assumption that dark matter is cold for the  purposes of galaxy formation implies that the primordial power spectrum has deviations over CDM at galactic scales.
This can happen for example in models such as warm dark matter (WDM), in which dark matter particles undergo free-streaming in the early Universe \citep{Colin2000,Bode2001}, as well as interacting dark matter, in which dark matter particles interact with relativistic particles in the early Universe \citep{Boehm2002,Buckley2014}.
The cutoff in the linear power spectrum reduces the severity of the missing satellite problem \citep{Boehm2014}, the too-big-to-fail problem and the dwarf rotation curve diversity problem \citep{ETHOS2,Zavala19v100}.
However, observations of the Ly-$\alpha$ forest constrain the mass of thermal WDM to $m_{\text{WDM}} \gtrsim 3.5$ keV (\citealt{Viel2013,Irsic2017}; although see \citealt{Garzilli2019}).
Although WDM also naturally predicts central density cores due to the upper bound on the phase space density set by the primordial thermal velocity dispersion \citep{Dalcanton01v561}, allowed WDM models cannot create large dark matter cores without severely under-predicting the abundance of low-mass galaxies \citep{Maccio2012}.
Similarly, the interacting dark matter model does not alleviate the cusp-core problem since the interactions between dark matter and the relativistic particles decouple long before the onset of dark matter haloes.

Alternatively, models in which dark matter particles interact with each other are known as Self-Interacting Dark Matter \citep[SIDM,][]{Spergel2000} models.
SIDM has self-interaction cross sections with an amplitude near that of strong nucleon-nucleon elastic scattering, which are sufficient to reduce the central densities of DM haloes and alleviate the CDM challenges \citep[for a review, see][]{Tulin18v730}. 
For example, by transferring energy from the outer regions of the halo to the inner regions, SIDM models can create cores on kpc scales \citep{Colin2002}. 
This alleviates both the cusp-core and the too-big-to-fail problems \citep{Vogelsberger2012, Rocha2013, Zavala2013}, and also the dwarf rotation curve diversity problem (\citealt{Kamada2017}; although see \citealt{Santos-Santos20v495}).

Most SIDM simulations thus far have assumed that the scattering process between two SIDM particles is elastic, i.e. kinetic energy is conserved during the collision.
In these purely elastic SIDM models, elastic scattering leads to a redistribution of dark matter particles within the halo, which has been found to modify DM haloes in terms of their phase-space structure \citep{Vogelsberger2013a} and halo shapes \citep{Peter13v430,Brinckmann2018}.
However self-scattering interactions can also be inelastic and occur in theoretical models which contain multi-state dark matter \citep {AH2009,Schutz2015}.
For the case of a two-state scenario, a transition from the ground ($\chi_1$) to the excited ($\chi_2$) state (up-scattering) is an endothermic process, while the reverse process (down-scattering) is exothermic. 
Inelastic SIDM is especially interesting from a core-formation perspective, since a down-scattering event produces a kick in the velocities of the ground state particles.
For example, \cite{Todoroki19v483,Todoroki19v483b,Todoroki2020} performed SIDM simulations using a multi-state dark matter model and concluded that elastic scattering and energy injection are independently sufficient to create isothermal cores in haloes of mass ${\approx} 5 \times 10^{11} M_\odot$.
The energy injected into the halo from exothermic down-scattering can be on the order of 100 million Type II supernovae, which not only results in higher core formation efficiencies than a purely elastic model, but also reduces the amount of substructure present in the halo \citep[][hereafter \citetalias{Vogelsberger19v484}]{Vogelsberger19v484}.
As such, inelastic SIDM can increase the allowed parameter space of self-interaction cross sections in SIDM models.
So far, the impact of energy injection due to inelastic self-interactions on the dark matter properties of galactic haloes remains largely unexplored.

In this paper, we examine the effects of inelastic SIDM on halo structure and assembly using  high-resolution DM-only simulations of a MW-size halo.
The study is based on a comparative analysis between  a CDM model, an elastic SIDM model, and an inelastic SIDM model.
By using DM-only simulations, we focus solely on the impact of SIDM in the absence of gas physics and galaxy formation.
The paper is structured as follows: 
Section 2 describes the numerical simulations, including a description of the two-state inelastic SIDM model and its implementation.
We then present the results of the simulations, focusing on the effects of inelastic SIDM on the structure of the simulated halo at $z=0$ in Section 3, as well as its assembly history in Section 4.
Finally, we present our conclusions in Section 5.

\section{Methods} 

Since the haloes analysed in this paper were previously introduced in \citetalias{Vogelsberger19v484}, we refer the reader there for a full  description of the model and methods.
We give a brief overview of the relevant points here.

\subsection{Inelastic SIDM model}
\label{model}
Our SIDM simulations are based on the two-state model presented by \cite{Schutz2015}, where the excited state is nearly degenerate with the ground state, and analytic expressions for the elastic and inelastic s-wave cross sections have been derived.
Although other inelastic SIDM models exist, this model is currently the only one in which an analytic description of scattering has proven to be feasible at dwarf galaxy velocity scales. 
The model is described by the following parameters: the mass splitting $\delta$ between the ground ($\chi^1$) and excited ($\chi^2$) states, the coupling constant $\alpha$ between the dark matter particle and the force mediator, the mass $m_{\chi}$ of the dark matter particle, and the mass $m_{\phi}$ of the force mediator. 
In this work, these parameters have the values $\delta=10$~keV, $\alpha=0.1$, $m_{\chi}=10$~GeV, and $m_{\phi}=30$~MeV.
While these values represent an arbitrary point within the entire parameter space, they have been chosen to lie within the range for interesting cross sections for velocities of the order of 10~\kms{} \citep{Schutz2015}. 
At the scale of the MW halo, the corresponding elastic cross section per unit mass is a few~$\text{cm}^2 \text{ g}^{-1}$.
Such cross sections have previously been found to be capable of creating cores of size $\mathcal{O}({\rm kpc})$ \citep[e.g.][]{Vogelsberger2012,Brinckmann2018}

There are five possible interactions in this two-state SIDM model, namely: 
\begin{enumerate}
    \item elastic scattering of two ground state particles $(\chi_1 + \chi_1 \rightarrow \chi_1 + \chi_1)$,
    \item elastic scattering of two excited state particles $(\chi_2 + \chi_2 \rightarrow \chi_2 + \chi_2)$,
    \item elastic Yukawa scattering $(\chi_1 + \chi_2 \rightarrow \chi_1 + \chi_2)$,
    \item inelastic endothermic up-scattering in which two ground state particles transition to the excited state $(\chi_1 + \chi_1 \rightarrow \chi_2 + \chi_2)$, and \item inelastic exothermic down-scattering in which two excited particles transition to the ground state $(\chi_2 + \chi_2 \rightarrow \chi_1 + \chi_1)$.
\end{enumerate}
During down-scattering, our chosen model produces a velocity kick $ v_{\rm kick} = \sqrt{2\delta/m_{\chi}} c \simeq 424$~\kms{}.
Up-scattering can only occur for relative velocities $v_{\rm rel} > 2 v_{\rm kick} \simeq 848$~\kms{}.
For typical dark matter velocities in a MW-size halo (${\approx} 200$~\kms{}), inelastic endothermic scattering is essentially forbidden since the energy splitting $\delta$ is too large for this interaction  to occur frequently.

An additional parameter which must be specified in the simulations is the primordial fraction of dark matter in each of the ground and excited states.
A primordial excitation fraction \chih{} corresponds to having all particles being in the excited state, and leads to the maximum possible energy release during structure formation and hence the maximum possible effect on halo structure.
Conversely, an inelastic system with all particles initially in the ground state behaves essentially like purely elastic SIDM, since inelastic up-scattering is suppressed at galactic halo velocities.
In this paper, we examine two initial configurations: 
(i) \chih{}, where all SIDM particles begin in the excited state, 
and (ii) \chif{}, where only half of the SIDM particles begin in the excited state.
For possible theoretical justifications for the large primordial fraction of excited states, we refer the reader to \citetalias{Vogelsberger19v484}.

\subsection{Numerical implementation}

The two-state inelastic SIDM model is implemented within a general multi-state dark matter framework in the {\sc Arepo} code \citep{Springel2010}. 
This framework is a generalisation of the probabilistic approach presented in \cite{Vogelsberger2012} and \cite{ETHOS2}, and is able to handle an arbitrary number of states with non-degenerate energy level splittings and all possible reactions between them, each with any given velocity-dependent cross section. 

Each dark matter particle $i$ is assumed to be in a specific state $\alpha$. 
The simulation volume is populated by dark matter particles in various states $(\alpha, \beta, \gamma, \delta)$ with possible two-body scatterings:
\begin{equation}
    \chi_i^{\alpha}+\chi_j^{\beta} \rightarrow \chi_i^{\gamma}+\chi_j^{\delta}.
\label{eqn:scattering}
\end{equation}
Equation~\ref{eqn:scattering} represents the scattering of particles $i$ and $j$ from states $\alpha$ and $\beta$ into states $\gamma$ and $\delta$. 
A dark matter particle in state $\epsilon \in (\alpha, \beta, \gamma, \delta)$ has mass $m_{\chi^{\epsilon}}$. 
The scattering rates for the possible reactions are given by:

\begin{equation}
    R^{\alpha\beta \rightarrow \gamma\delta} = \frac{\rho^{\beta}}{m_{\chi^{\beta}}} \langle \sigma_T^{\alpha\beta \rightarrow \gamma\delta}(v^{\alpha\beta})v^{\alpha\beta} \rangle,
    \label{eqn:scatteringrate}
\end{equation}
where $\rho^{\beta}$ is the local mass density of particles in state $\beta$, $\sigma_T^{\alpha\beta \rightarrow \gamma\delta}(v^{\alpha\beta})$ is the velocity-dependent transfer cross section for the reaction, and $v^{\alpha\beta}$ is the magnitude of the relative velocity between the interacting particles.
Since the differential cross section has no angular dependence in this model, the transfer cross section is also the same as the total cross section for the reaction. 

The scattering of particles $i$ and $j$ in states $\alpha$ and $\beta$ with masses $m_i^{\alpha}$ and $m_j^{\beta}$ is performed in the centre of mass frame, and new velocities are assigned to each particle:
\begin{align} 
\mathbf{v}_i = \frac{m_i^{\alpha}+m_j^{\beta}}{m_i^{\gamma}+m_j^{\delta}} \textbf{v}_{\text{cm}} + \frac{m_j^{\delta}}{m_i^{\gamma}+m_j^{\delta}} A_{ij} v_{ij} \hat{\textbf{e}},\\
\textbf{v}_j = \frac{m_i^{\alpha}+m_j^{\beta}}{m_i^{\gamma}+m_j^{\delta}} \textbf{v}_{\text{cm}} - \frac{m_i^{\gamma}}{m_i^{\gamma}+m_j^{\delta}} A_{ij} v_{ij} \hat{\textbf{e}},
\label{eqn:sidm_vel}
\end{align}
%\tilde{v}_{ij}^{\alpha\beta \rightarrow \gamma\delta}
where $\textbf{v}_{\text{cm}}$ is the velocity of the centre of mass, $\hat{\textbf{e}}$ is a random vector on the unit sphere, and $A_{ij}(\alpha\beta \rightarrow \gamma\delta)$ is a dimensionless velocity scale factor that depends on the energy splitting of the reaction:
\begin{equation}
0 \leq A_{ij} \begin{cases} 
      =1 & \text{elastic} \\
      <1 & \text{inelastic: endothermic} \\
      >1 & \text{inelastic: exothermic}
   \end{cases}
   \label{eqn:aij}
\end{equation}
For inelastic scattering, each particle in the reaction gains or loses the same amount of energy. 
The endothermic case has as a lower limit the completely inelastic collision, in which both particles move at the centre of mass velocity. 
On the other hand, the exothermic case has no upper limit - 
the amount of energy injected into the system in the exothermic case is determined by the mass splitting $\delta$ between the two states.

\subsection{Numerical simulations}

We perform five high resolution DM-only simulations of a single MW-size halo in a cosmological context.
In addition to simulating the halo with the inelastic SIDM model, simulations with an elastic SIDM model as well as standard CDM are also performed for comparison.
For each SIDM model, we examine cases with the two primordial excited fractions \chih{}, and \chif{}.

The elastic SIDM model we examine is obtained by neglecting energy changes for the inelastic up and down-scattering reactions i.e. keeping $A_{ij}=1$ in Equation \ref{eqn:sidm_vel}.
The resulting elastic SIDM model is otherwise identical to inelastic SIDM, which enables us to isolate the effects of energy injection into the halo.

Our simulations used cosmological parameters consistent with Planck \citep[$\Omega_m$~=~0.302, $\Omega_\Lambda$~=~0.698, $\Omega_b$~=~0.046, $h$~=~0.69, $\sigma_8$~=~0.839, $n_s$~=~0.967,][]{Planck2014,Spergel2015}. 
The gravitational softening length is fixed in comoving coordinates until $z=9$ and then fixed in physical units until $z=0$, resulting in a Plummer-equivalent softening length of 72.4~pc at $z=0$.
The dark matter particles have a mass resolution of $2.756 \times 10^4 M_\odot$.
Haloes are identified using a friends-of-friends ({\sc fof}) algorithm with a linking length of 0.2 \citep{Davis85}, and the \textsc{subfind} algorithm is subsequently used to identify gravitationally self-bound subhaloes \citep{Springel01v328,Dolag09v399}.
In the CDM simulation, the virial mass
of the halo at $z=0$ is $M_{\text{200}} = 1.6 \times 10^{12} M_{\odot}$ and the virial radius is $R_\text{200} = 243$~kpc\footnote{
$R_\text{200}$ is the radius within which  the average density is $\bar \rho = 200 \rho_\text{crit}$, where $\rho_\text{crit}$ is the critical density of the universe. 
Other properties for the CDM and SIDM haloes can be found in Table 1 of \citetalias{Vogelsberger19v484}.}.

\section{Halo Structure}

In this section, we focus on the effect of inelastic SIDM on the structure of the dark matter halo at $z=0$, particularly in terms of the
density profile, velocity profile, phase-space structure, and halo shape.

\subsection{Density profile}
\label{sec:density}

\begin{figure*}
	\includegraphics[width=0.95\textwidth]{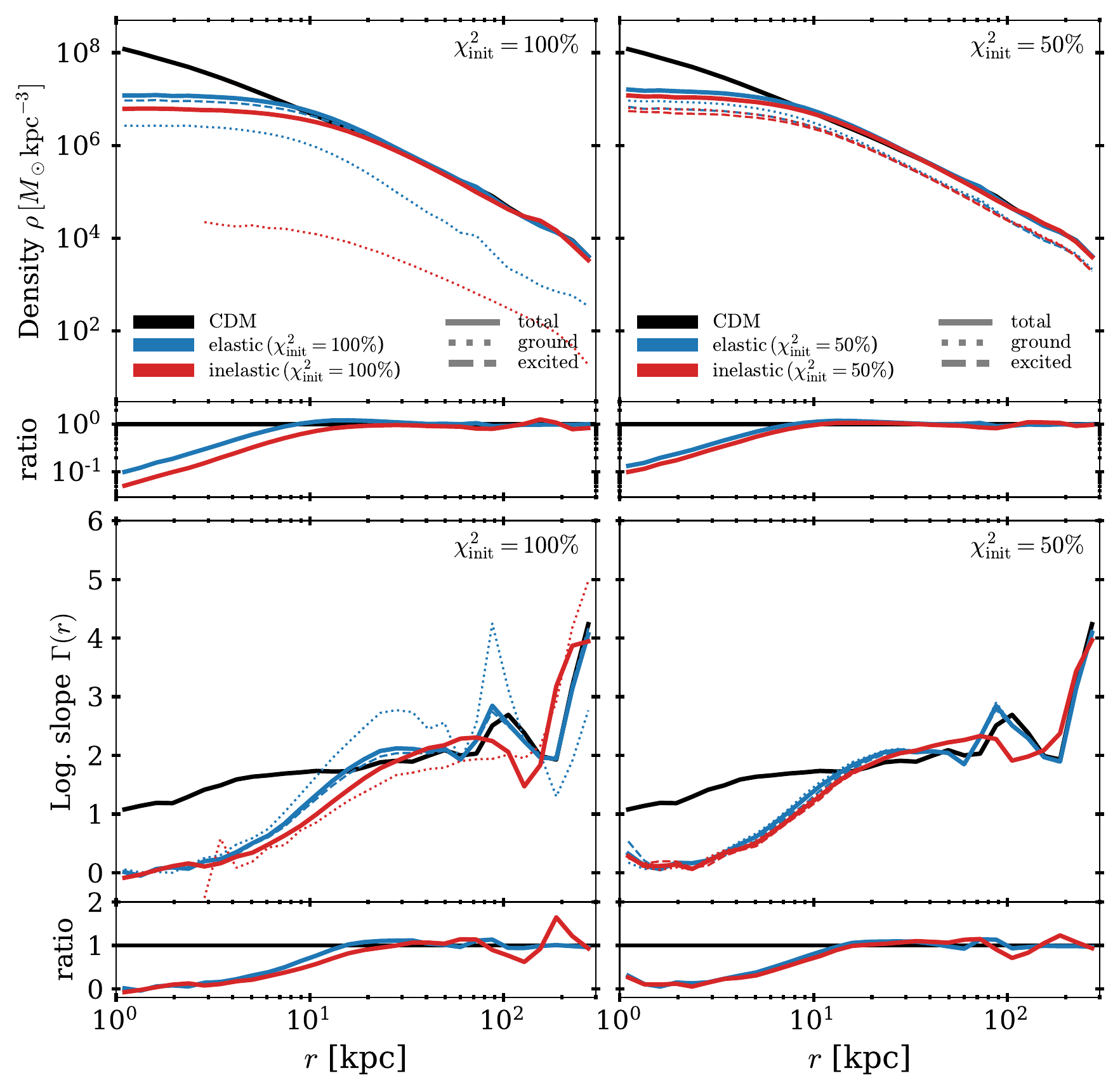}
    \caption{\textbf{Spherically averaged radial density profile (\emph{top panels}) and its logarithmic slope (\emph{bottom panels}) for the Milky Way-size halo.} 
    The columns are for different primordial fractions of the excited states, \chih{} (left) and \chif{} (right).
    We show the results for CDM (black), elastic SIDM (blue) and inelastic SIDM (red), with the narrow plot at the bottom of each panel showing the ratio of each quantity relative to CDM.
    The dotted (dashed) lines show the profiles for only the ground (excited) state.
    \emph{Top panels}: In the central regions, both SIDM models lead to the formation of density cores, with central densities depleted compared to CDM.
    Inelastic SIDM produces larger and lower density cores than elastic SIDM, since the exothermic down-scattering reactions lead to an expulsion of ground state particles from the halo.
    At intermediate radii, the elastic density profile exceeds that of CDM while the inelastic profile almost never does so. 
    This is because particles are only redistributed from the central regions to the intermediate regions in the elastic case and are not removed as in the inelastic case.
    For the excitation fraction \chih{}, the removal of down-scattered ground state particles leads to a significant suppression of the ground state inelastic density profile (blue dotted lines) relative to elastic SIDM.
    \emph{Bottom panels}: Radial profiles of the logarithmic density slope $\Gamma(r)= - d\ln\rho / d\ln r$.
    In all SIDM cases, the slope reaches a value of zero, indicating clearly the presence of density cores.
    }
    \label{fig:denslope}
\end{figure*}

We begin by examining the spherically-averaged DM density profile $\rho(r)$ of the MW-size halo.
The particles are binned into 30 shells spaced in logarithmic intervals in the range 1~kpc < $r$ < 300~kpc.
The results are shown in the upper panels of Fig. \ref{fig:denslope}, with black, blue, and red lines representing the CDM, elastic SIDM, and inelastic SIDM models, respectively. 
The columns distinguish between the primordial excitation fractions \chih{} (left panels), and \chif{} (right panels).
In addition to the total mass density (solid lines), we also show the individual profiles for the ground (dotted lines) and excited states (dashed lines).
To avoid shells with low number counts, shells containing less than 20 particles are ignored.
This affects primarily the inelastic model with \chih{} (left panel, dotted red lines).
We note that the density profiles were previously also presented in Fig.~8 of \citetalias{Vogelsberger19v484}.

The upper row of Fig. \ref{fig:denslope} shows that both elastic and inelastic SIDM models produce lower densities and extended cores near the halo centre compared to CDM.
At $r = 1$~kpc, elastic SIDM reduces the total density by a factor of $\sim$10 relative to CDM.
Defining the core size $R_\text{core}$ as the radius below which the SIDM density profiles deviate from CDM, we estimate the elastic SIDM haloes to have density cores of size $R_\text{core} \sim9$~kpc.

Inelastic self-interactions further increase the core size and decrease the central density compared to elastic SIDM.
For the \chih{} excitation fraction (upper left panel), the total density at $r = 1$~kpc is approximately $20$ times smaller than that of the CDM halo, with a core size of $R_\text{core} \sim 20$~kpc.
Furthermore, the ground state densities (dotted lines) exhibit a large difference between the inelastic and elastic SIDM models: 
the ground state density of the inelastic halo is significantly lower compared than that of the elastic counterpart, with a difference of approximately two orders of magnitude near the halo centre.
For the lower excitation fraction \chif{}, the ground state density is only slightly above that of the excited state, since the de-excited particles only constitute a small fraction of the total ground state population.

For each SIDM case, we further plot the ratio of the SIDM profiles compared to CDM ($\rho_{\rm SIDM}/\rho_{\rm CDM}$) in the narrow plot at the bottom of each panel.
At intermediate radii (10~kpc~$\gtrsim~r~\gtrsim$~100~kpc), we note that the density of the inelastic SIDM haloes actually exceed that of the CDM counterpart.
This is due to the redistribution of particles from the inner regions during the  elastic scattering process. 
However, the particles remain bound to the halo in the absence of an energy-injection mechanism.
In contrast, the density profile of the inelastic \chih{} case does not exceed that of CDM at the same radii.
Here, the down-scattering process imparts velocity kicks, which enable the de-excited ground state particles to travel further and even escape from the inelastic halo completely.
As a result, in addition to a redistribution of particles, there is a removal of particles from the halo (\emph{halo evaporation}) which decreases its overall density and equivalently, mass.
These velocity kicks inject energy equivalent to hundreds of millions of SNII into the halo \citepalias{Vogelsberger19v484}, which further explains the more efficient core formation with the inelastic SIDM model.
The slight elevation of the inelastic density profile at $r \approx 130$~kpc compared to CDM, also reflects down-scattered particles on their way out towards the outskirts of the halo.

The bottom panels of Fig.~\ref{fig:denslope} show the  slopes $\Gamma(r) \equiv -d\ln{\rho} / d\ln{r}$ of the logarithmic density profiles. 
The slopes are calculated from the density profiles $\rho(r)$ using a central difference scheme, such that for the $n$-th bin, we have
\begin{equation}
    \Gamma(r_n)= -\frac{ \ln(\rho(r_{n+1}))-\ln(\rho(r_{n-1})) }{ \ln(r_{n+1})-\ln(r_{n-1}) }.
\end{equation}
For the first and last radial bins, forward and backward schemes are used respectively to estimate the slopes.
In the CDM case, the Navarro-Frenk-White (NFW) profile \citep{Navarro1996}, with a cusp ($\Gamma \approx 1$) at small radii and $\Gamma \approx 3$ near the virial radius, is a good fit to the halo density profile.
On the other hand, in all SIDM models, we find that at small radii (1~kpc < $r \lesssim$ 3~kpc), the central slope is $\Gamma \approx 0$, revealing the presence of density cores.
The overall slopes of the inelastic models beyond the density cores remain below that of the elastic model, up to intermediate radii of $r \lesssim 20$~kpc.
This effect is stronger and persists to a larger radius for the larger primordial excitation fraction.

Another key difference between the elastic and inelastic models lies in the slopes of the ground state particles for \chih{} (lower left panel).
In the elastic model, the ground state density profile is steeper (larger $\Gamma$) compared to the excited state beyond the density core.
The opposite trend occurs for the inelastic model, where the ground state density profile is shallower (smaller $\Gamma$) than that of the excited state.

In general, it is clear that the impact of inelastic SIDM depends strongly on the initial fraction of excited particles: a higher primordial excited  fraction results in larger modifications to the density profile, a result of the larger amount of energy injected into the halo.

\subsection{Velocity dispersion and anisotropy}
\label{sec:veldisp}

\begin{figure*}
	\includegraphics[width=0.95\textwidth]{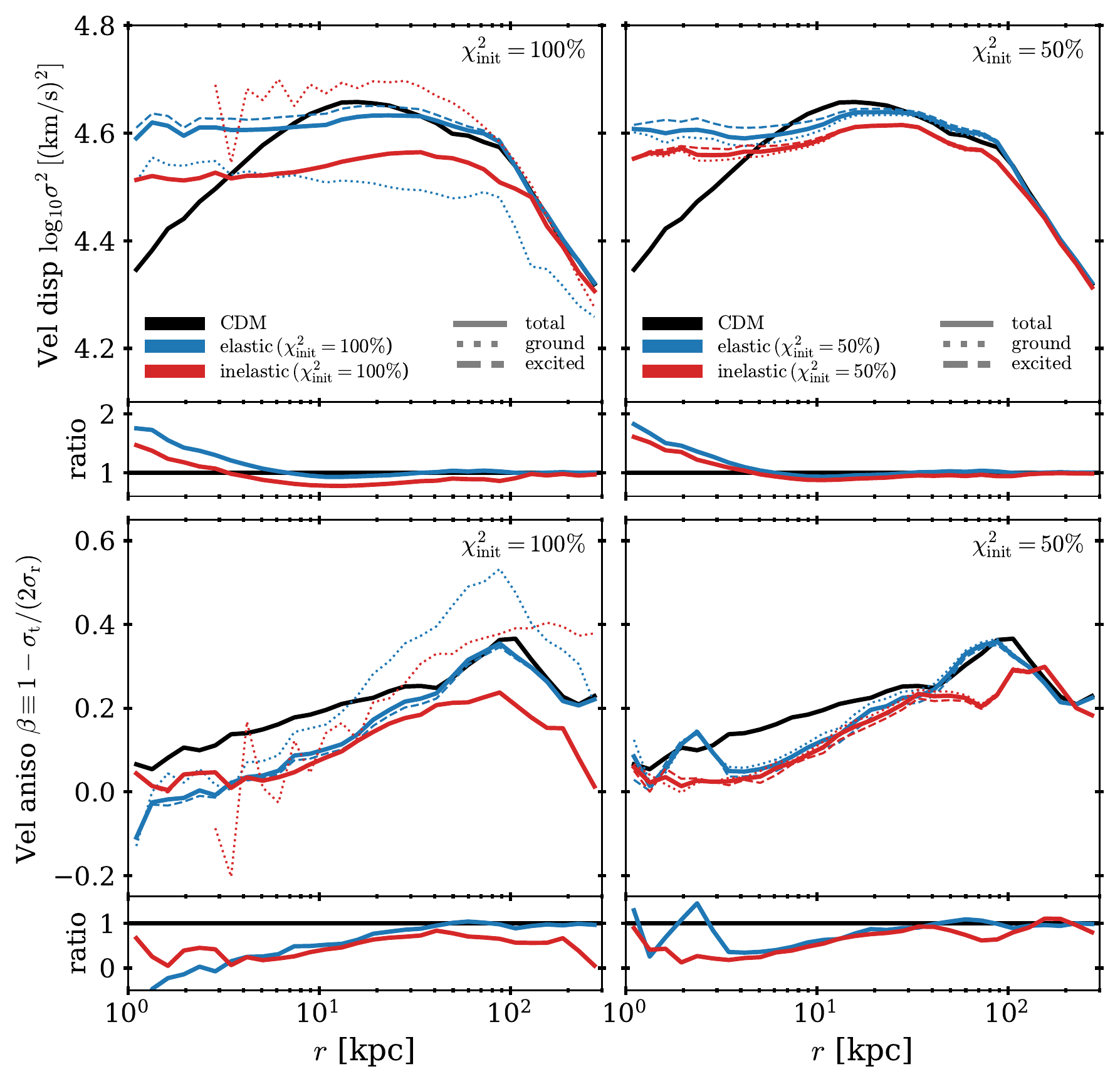}
    \caption{\textbf{Velocity dispersion (top panels) and velocity anisotropy profiles (bottom panels) for the Milky Way-size halo}.
    The narrow plot at the bottom of each panel shows the ratio of the SIDM quantity relative to CDM.
    \emph{Top panels}: The SIDM velocity dispersion profiles are flat (isothermal) in the inner regions, exceeding that of CDM which exhibits the well-known temperature inversion.
    Inelastic scattering results in a suppression of the velocity dispersion compared to elastic SIDM, a difference which is most apparent with a primordial excited fraction of \chih{}.
    \emph{Bottom panels}: The velocity anisotropy $\beta$ of the SIDM haloes is closer to $\beta = 0 $ (more isotropic) compared to the CDM counterpart.
    Although the difference between SIDM and CDM decreases in general towards the virial radius, the inelastic SIDM halo with excitation fraction \chih{} remains significantly more isotropic than the CDM halo even at the virial radius.
    %This difference decreases towards the virial radius, at which point only the inelastic SIDM model with \chih{} maintains a smaller velocity anisotropy. 
    }
    \label{fig:veldisp}
\end{figure*}

\begin{figure*}
	\includegraphics[width=\textwidth]{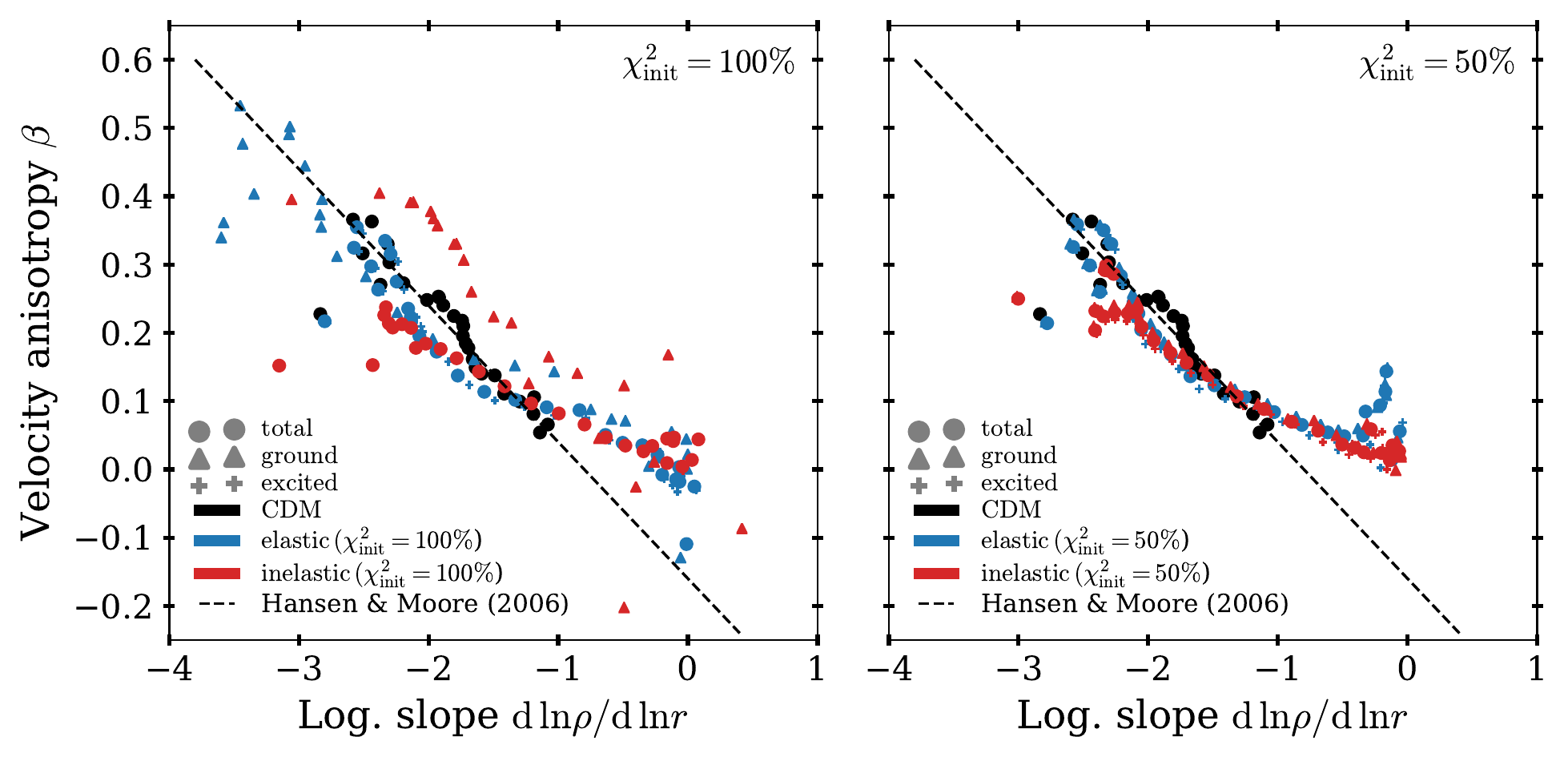}
    \caption{\textbf{Velocity anisotropy parameter vs logarithmic slope of the density profile for the Milky Way-size halo.} 
    The dashed black line represents the linear relation proposed by Hansen \& Moore (2006).%\cite{Hansen2006}. 
    The CDM halo agrees with the \citetalias{Hansen2006} relation whereas the SIDM haloes diverge from this linear relation.
    The difference is particularly significant for $d\ln\rho/d\ln r \gtrsim -1$, which corresponds to the core region in the SIDM haloes.
    In this region, a simple extrapolation of the CDM result leads to an under-prediction of the velocity anisotropy since the flattening of the density profile is not taken into account.
    }
    \label{fig:GB14}
\end{figure*}

The velocity dispersion and anisotropy are useful in providing insights into the orbital structure of haloes since the collision of dark matter particles can affect their velocity distributions due to the thermalisation of the halo \citep{Vogelsberger2013a,Brinckmann2018}. 
We calculate the velocity dispersion $\sigma^2$ as
\begin{equation}
    \sigma^2(r) = \left<\left| \mathbf{v} - \bar{\mathbf{v}} \right|^2 \right> 
    = \frac{1}{N_p} \sum_{i=1}^{N_p} \left| \mathbf{v}_i - \bar{\mathbf{v}} \right|^2
    \label{eqn:veldisp}
\end{equation}
where $N_p$ is the number of particles and $\bar{\mathbf{v}}$ is the mean velocity of particles in a given spherical shell. The velocity anisotropy parameter is defined as
\begin{equation}
    \beta (r) =  1 - \frac{\sigma_t^2(r)}{2\sigma_r^2(r)}
\end{equation}
where $\sigma_t^2$ and $\sigma_r^2$ are the tangential and radial velocity dispersions averaged over spherical shells, calculated similarly to Equation~\ref{eqn:veldisp}. 
Haloes with $\beta = 0$ are considered isotropic, while $\beta > 0$ and $\beta < 0$ correspond to radially and tangentially biased haloes respectively.
We ignore substructure in calculating the velocity anisotropy in order to concentrate on the smooth component of the halo.

Fig. \ref{fig:veldisp} shows both the radial profiles of the velocity dispersion $\sigma^2 (r)$  (top row) and the velocity anisotropy parameter $\beta(r)$ (bottom row), with the columns distinguishing between the two primordial excitation fractions.
The velocity dispersion of the CDM halo exhibits the well-known temperature inversion in the central region, where the velocity dispersion drops towards the centre, consistent with previous CDM simulations \cite[e.g.][]{Navarro2010,Tissera10v406}.
By transporting energy into the centre, self-interactions substantially affect the dark matter velocities, erasing the temperature inversion in the SIDM models.
Thus, the SIDM velocity dispersions flatten and become fairly constant (isothermal) towards the inner halo.
Inelastic down-scattering further reduces the velocity dispersions compared to the elastic counterparts, since kinetic energy is carried away by escaping down-scattered particles. 
At intermediate radii, the velocity dispersion of  the inelastic \chih{} halo (red curve, left panel) is $\approx$10 per cent lower than the CDM counterpart, a difference which persists for a substantial fraction of the halo, up to $r \approx 100$~kpc.

An additional impact of inelastic self-interaction can also be observed from the ground state velocity dispersions.
For the higher primordial excitation fraction \chih{} (Fig.~\ref{eqn:veldisp}, left panel), velocity kicks received by the ground state particles in the inelastic SIDM model increase the inelastic velocity dispersion of the ground state (dotted line) over that of the excited state particles (dashed line).
Conversely, for elastic SIDM, the velocity dispersion of ground state particles is lower than of the excited state.
In this case, ground state particles are only ever the products of elastic interactions, and are therefore more limited in their allowed velocities.

The bottom panels of Fig.~\ref{fig:veldisp} present the radial profiles of the velocity anisotropy parameter $\beta (r)$.
In general, self-interactions isotropize the orbits ($\beta$ closer to zero) compared to CDM, especially in the inner and intermediate regions of the halo.
Near the halo centre, $\beta$ approaches zero more rapidly in the SIDM models than in CDM, due to collisional relaxation driving the core region isothermal \citep{Colin2002}.
Whereas elastic SIDM differs from CDM only in the inner and intermediate regions, the effects of inelastic SIDM persist up to and even beyond the virial radius, especially for the \chih{} excitation fraction.
This result confirms once again that energy injection from inelastic down-scattering plays an important role in the structure of the halo far from the halo centre.

%The velocity anisotropy of the down-scattered ground state particles can also be observed from the bottom-left panel of Fig.~\ref{fig:veldisp}. 
%At intermediate to large radii, we find that the down-scattering processes increases the velocity anisotropy of the ground state particles relative to the excited state.

The effect of SIDM on halo structure is also evident when plotting the anisotropy parameter $\beta$ against the logarithmic slope of the density profile ($d\ln\rho/d\ln r \equiv -\Gamma$), shown in Fig. \ref{fig:GB14}.
A relation between these two quantities were first identified by \cite{Hansen2006} (hereafter \citetalias{Hansen2006}) in CDM simulations.
This relation is shown as black dashed lines in Fig. \ref{fig:GB14}.
%Using CDM simulations,  identified a relation between the density slope and velocity anisotropy within a halo, shown as black dashed lines in.
Our CDM results (black dots) is in good agreement with the \citetalias{Hansen2006} relation, which predicts a central velocity anisotropy and slope of $(\beta(0),-\Gamma(0)) = (0,-1)$.

In the following, we focus on results that include both DM states (i.e. the filled circles).
For steep slopes ($d\ln\rho/d\ln r \lesssim -1$), the SIDM results are close to, but lie slightly below the \citetalias{Hansen2006} relation.
This is due to the increased velocity isotropy (smaller $\beta$) at intermediate radii in the SIDM runs.
In the core region, ($d\ln\rho/d\ln r \gtrsim -1$), a simple extrapolation of the \citetalias{Hansen2006} relation under-predicts the velocity anisotropy compared to the SIDM run since the CDM results do not account for the flattening of the density profile.
The 
Contrary to CDM, both SIDM models predict a central velocity anisotropy and slope of $(\beta(0),-\Gamma(0)) \approx (0,0)$, regardless of the primordial  fraction of excited states.

\subsection{Phase-space structure}

\begin{figure*}
    \centering
	\includegraphics[width=\textwidth]{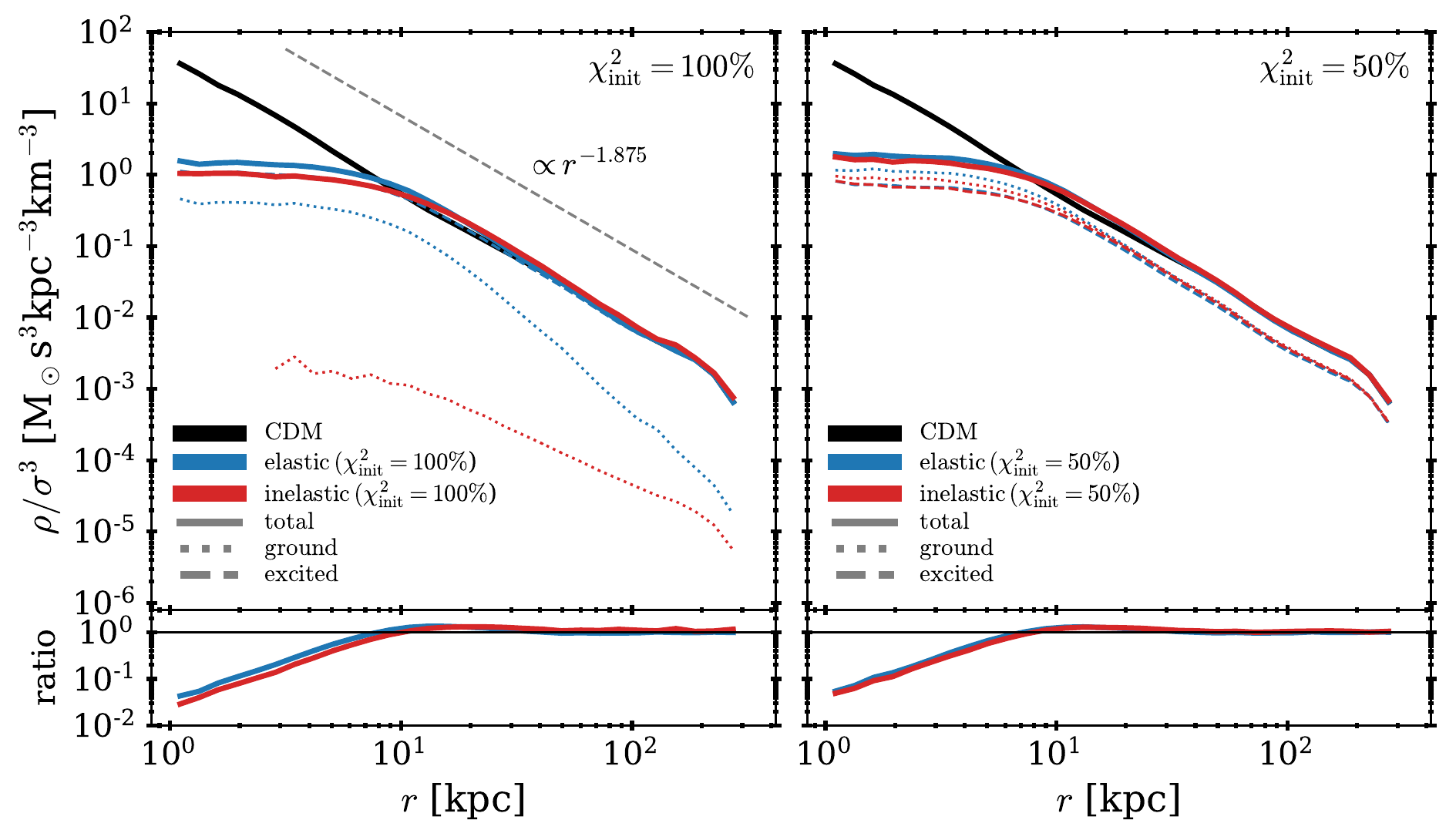}
    \caption{{\bf Pseudo phase-space density as a function of radius for the Milky Way-size halo.} 
    Whereas the CDM profile agrees with the power law profile ($\rho/\sigma^3 \propto r^{-1.875}$) found by Taylor \& Navarro (2001), both elastic and inelastic SIDM lead to a flattening of the pseudo phase-space density profile towards the centre, due to a combination of core formation and flattening of the velocity dispersion.
    }
    \label{fig:phasespace}
\end{figure*}

\begin{figure*}
    \centering
	\includegraphics[width=0.95\textwidth]{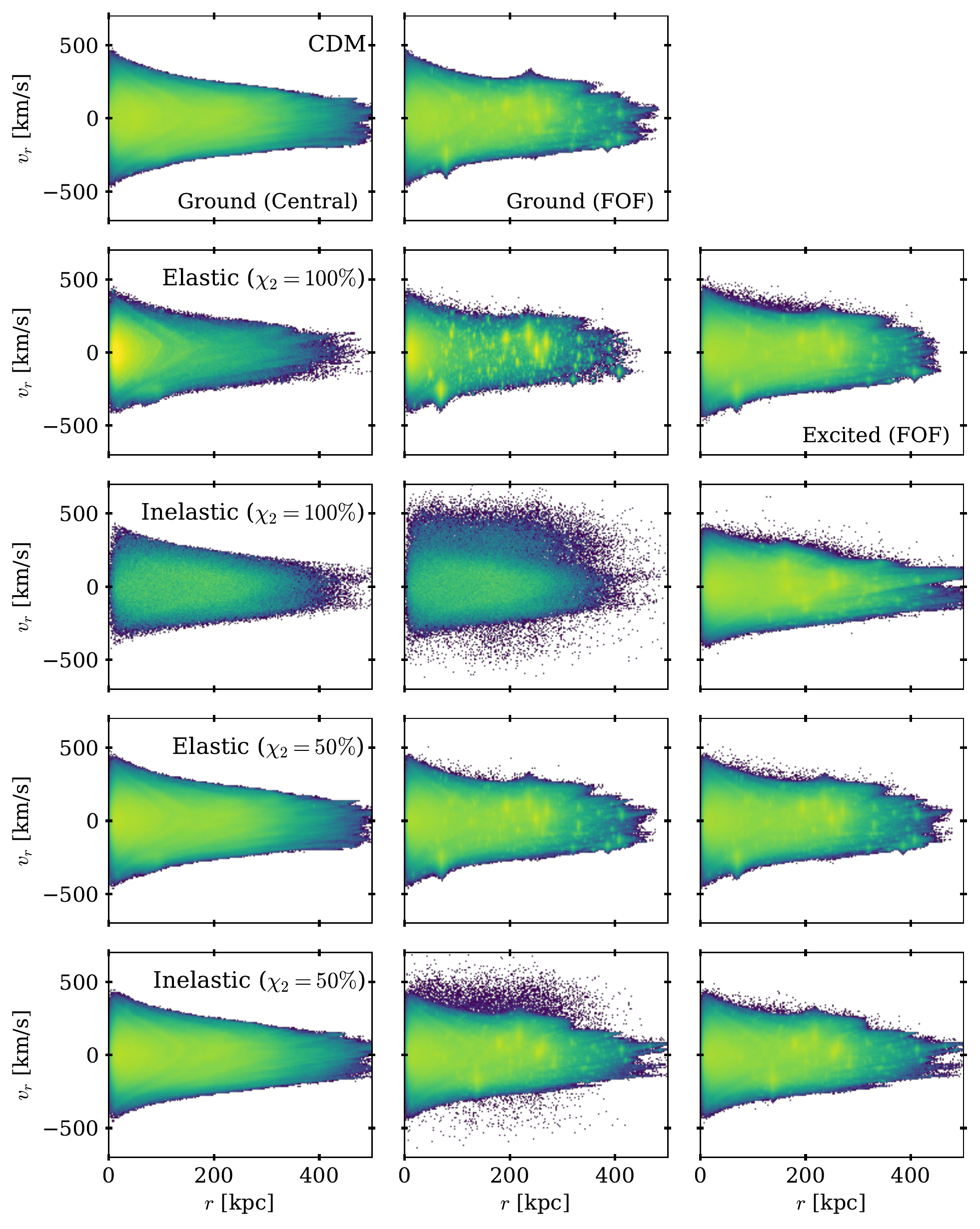}
    \caption{{\bf Radial phase-space histograms of all models for the Milky Way-size halo.}
    The pixel colour represents the particle count in each $(r,v_r)$ bin, with dark points representing low counts, enhancements representing high particle counts and uncoloured portions representing unoccupied regions of phase-space.
    We distinguish between  bound ground state particles in the main subhalo (leftmost column), all ground state particles within the {\sc fof} halo (middle column), and excited state particles within the {\sc fof} halo (rightmost column).
    Results for the {\sc fof} halo consider all substructures as well as unbound particles.
    Inelastic SIDM (third and fifth rows) results in a  population of ground state particles with radial speeds up to 500~\kms{} throughout the halo.
    For a primordial excitation fraction of \chih{}, the inelastic SIDM model strongly suppresses substructure in the ground state (third row, middle column).
    }
    \label{fig:phasespacehist}
\end{figure*}

In CDM $N$-body simulations, \cite{Taylor2001} discovered that the dark matter density and velocity dispersion of any halo can be combined to form a \emph{pseudo phase-space density}  $Q(r) \equiv \rho(r)/\sigma^3(r)$, a quantity which is inversely proportional to the local entropy. $Q(r)$ is essentially universal for all haloes, and can be approximated by a power law $Q(r) \propto r^{-\alpha}$. 
The value of $\alpha$ obtained ($\alpha \approx 1.875$) was also predicted by the analytic spherical infall solution of \cite{Bertschinger1985}.

In Fig. \ref{fig:phasespace}, we present the pseudo phase-space density profiles $Q(r)$ of the main halo for the CDM and SIDM models.
The CDM case obeys and confirms the power law seen in previous $N$-body simulations \citep[e.g.][]{Taylor2001,Ludlow2011}.
With SIDM however, the power-law behaviour is broken since self-scattering leads to a flattening in $Q(r)$ for $r < 10$~kpc.
This is due to the central flattening of the individual density and velocity profiles by self-interactions discussed in the previous sections.
Inelastic interactions further modify the pseudo phase-space profile, resulting in a decrease in $Q(r)$ in the inner halo compared to the elastic case, most notably for the higher primordial excitation fraction \chih{}.

To further illustrate the differences in the SIDM models, we plot the radial phase-space distributions in Fig.~\ref{fig:phasespacehist}.
Here, the pixel colour represents the particle count in each $(r,v_r)$ bin, with dark points representing low counts, enhancements representing high particle counts and uncoloured portions representing unoccupied regions of phase-space.
For ground state particles, we distinguish between particles bound to the central subhalo (leftmost column, as identified by {\sc subfind}), and all particles within the {\sc fof} group (middle column), which includes both unbound particles as well as substructure.
In Fig.~\ref{fig:phasespacehist}, CDM particles are considered as ground state particles.
For the SIDM models, we include also the phase-space distribution of all excited state particles in the {\sc fof} group (rightmost column).
%We use the same normalisation across all the histograms in  Fig. \ref{fig:phasespacehist}.

We first note that for a primordial excitation fraction of \chih{}, the phase-space distributions show a clear deficit of the ground state abundances in the inelastic SIDM model (third row) compared to the elastic model (second row), which agrees with our previous results for the halo density profiles (Fig.~\ref{fig:denslope}).
In the radial phase-space distributions, satellite subhaloes appear as enhancements in the number counts localised at particular radii and radial velocities.
The deficiency of ground state particles in the inelastic  \chih{} case carries over to the subhalo structure, which is noticeably absent in the ground state phase-space histogram of the {\sc fof} halo (third row, middle column).
Due to the small potential wells of low-mass satellite subhaloes, the ground state particles formed from exothermic down-scattering can easily escape due to the imparted velocity kicks. 
In the inelastic model, we find most of the resolved subhaloes\footnote{We consider a subhalo to be resolved if it contains at least 100 particles in total.} do not contain any ground state particles: only 49 out of 3304 resolved subhaloes have a non-zero ground state fraction.
In comparison, most of the subhaloes (4298 out of 4306) in the elastic model with the same primordial excitation fraction host ground state particles.
It is also important to note that subhaloes might not correspond exactly between the various runs for two reasons:
(i) inelastic self-interactions lead to the evaporation of subhaloes and thus a reduction in subhalo abundance \citepalias{Vogelsberger19v484}, and 
(ii) the temporal evolution of the counterparts deviate progressively as time goes by, since subhalo accretion histories and orbits are affected by dynamical changes generated by self-interactions. 

For both inelastic SIDM cases, Fig.~\ref{fig:phasespacehist} shows a population of ground state particles moving outwards with high radial velocities of up to ${\approx} 500$~\kms{}.
These high-speed ground state particles, noticeably absent in CDM and  the elastic SIDM model, reflect the velocity kicks they received during inelastic down-scattering.
Although the majority of these kicks result in outward-moving ($v_r > 0$) particles, they can occasionally  result in inward-moving ($v_r < 0$) particles, as evidenced from the smaller population of particles with high-speed negative radial velocities in the inelastic ground state phase-space distributions (third and bottom rows).
For the inelastic \chih{} case, we find that 80 per cent of these inward moving particles originate from accreted subhaloes, and the remaining one-fifth from the central halo.

%Subhaloes which have zero ground state fraction:
%Model_A_elastic: 8/4306
%Model_A_inelastic: 3255/3304
%Model_B_elastic: 0/4393
%Model_B_inelastic: 0/3986

\subsection{Velocity distribution at the solar circle}
\label{sec:solarvel}

\begin{figure*}
    \centering
	\includegraphics[width=\textwidth]{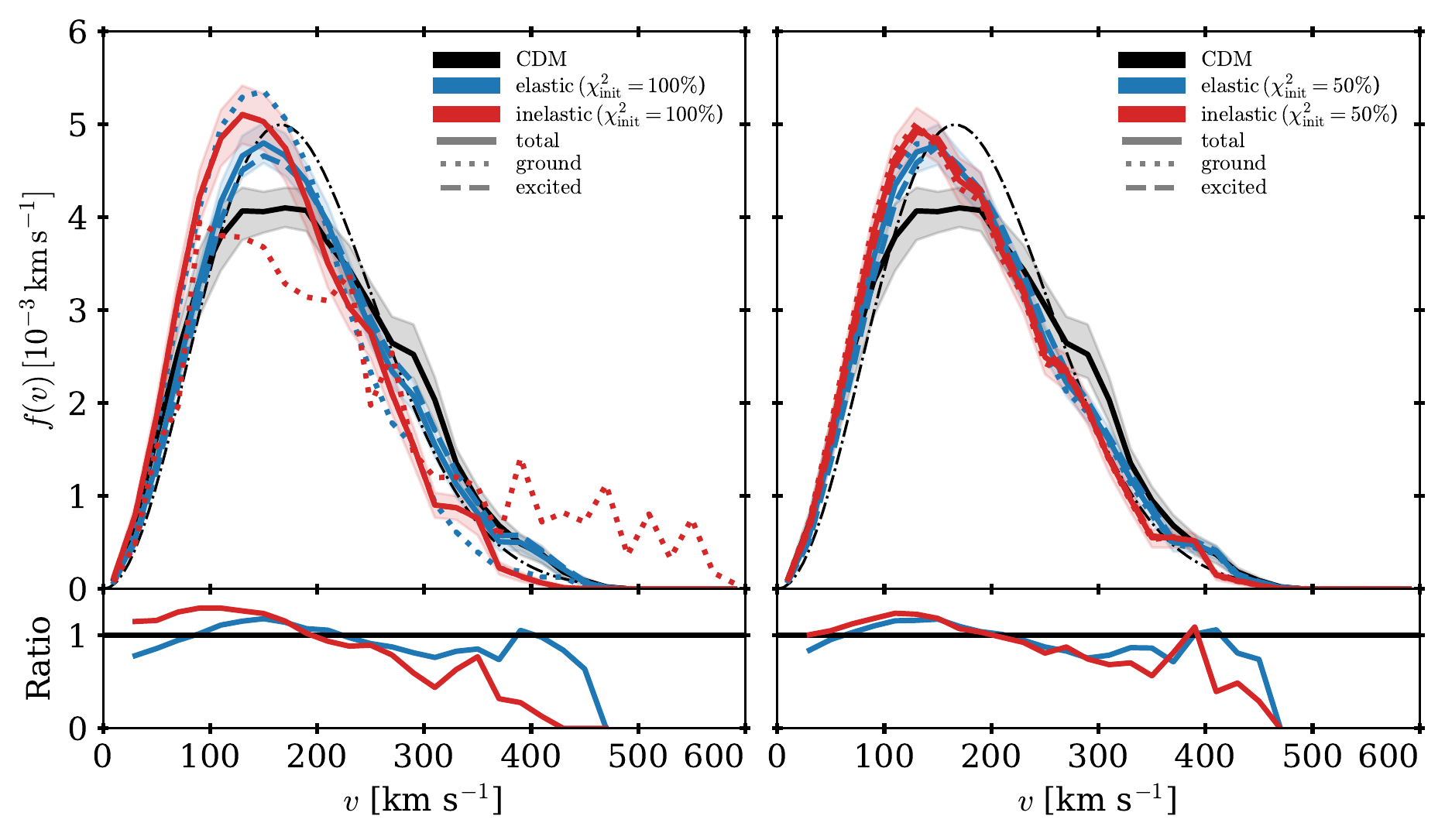}
    \caption{\textbf{Top panels: Local dark matter velocity distribution for the Milky Way-size halo at the solar circle.} 
    We measure $f(v)$ within cubes of side-length 2~kpc, sampled at a distance of $r = 8$~kpc for 1000 randomly selected observers.
    For each cube, we calculate the histogram in $v$ of all particles (ground + excited states), as well as for the ground and excited states separately.
    The curves show the median distribution, while the shading denotes the 25th-75th percentile across the 1000 random observers.
    The dotted-dashed line in each panel represents a Maxwellian distribution with mean velocity 187.3~\kms{}.
    At the bottom of each panel, we show the ratio of each SIDM case relative to CDM.
    Compared to CDM, the SIDM models decrease the overall peak velocity of the distribution and suppress the fraction of particles with velocities $v>200$~\kms{}.
    In the SIDM models, self-interactions cause the velocity distribution to become closer to Maxwellian.
    In the inelastic model with primordial excitation fraction \chih{} (left panel), the down-scattering of particles into the ground state results in population of high-speed ground state particles with speeds $v \gtrsim 400$~\kms{}. 
    Such a feature is absent in the velocity distribution of the excited particles and for the elastic SIDM model.
    }
    \label{fig:solarvel}
\end{figure*}

\begin{table}
    \begin{tabular*}{0.47\textwidth}{@{\extracolsep{\fill}}l c c}
    %\begin{tabular}{l c c}
    \hline
    Model &  mean $v$ & median $v$\\
          &    [\kms{}] &  [\kms{}]\\
    \hline
    CDM     & 187.3 & 179.4\\
    elastic SIDM (\chih) & 185.9 & 176.0\\
    elastic SIDM (\chif) & 183.3 & 172.7\\
    inelastic SIDM (\chih) & 168.5 & 159.4 \\
    inelastic SIDM (\chif) & 177.1 & 166.7\\
    \hline
    \end{tabular*}
    \caption{Mean and median of the (total) DM velocity distributions shown in Fig.~\ref{fig:solarvel}.}
    \label{tab:solarvel}
\end{table}

The local distribution of DM velocities is important for DM direct detection experiments, which depend on the shape of the local velocity distribution $f(v)$. 
%Here the quantity of interest is the local DM velocity distribution near the Earth.
%We measure the spherically-averaged velocity distribution at the solar circle i.e. $R_{\sun} = 8$~kpc 
We measure $f(v)$ within cubes of side-length 2~kpc, sampled at a distance of $r = 8$~kpc (the solar circle) for 1000 randomly selected observers\footnote{The cubes are oriented aligned with the simulation box. Changing the orientation e.g. by rotating the cubes is not found to affect the overall result.}.
%At this fixed halocentric distance, we calculate $f(v)$ within 1000 cubes of side-length 2~kpc randomly distributed over the shell
For each cube, we calculate the histogram in $v$ for all particles (ground + excited states), as well as the individual particle states, normalising each histogram such that $\int f(v) dv =1$.
The resulting velocity distributions are shown in Fig.~\ref{fig:solarvel}, with lines representing the median distributions over the 1000 random observers, and shaded regions denoting the 25th-75th percentile for the total (ground + excited state) distribution.
For comparison, we plot also a Maxwellian distribution (dotted-dashed curve) with mean speed equal to that of the CDM halo (187.3~\kms{}).
Table~\ref{tab:solarvel} reports the mean and median speeds of the total distributions for the five cases.

For CDM, our results are similar to previous work, where the velocity distribution is close to Maxwellian, with features that can be traced to the halo assembly history, as well as a tail of high-speed particles in excess of the best-fit Maxwellian distribution \citep[e.g.][]{Vogelsberger09v395,Kuhlen2010,Pillepich2014,Butsky16v462}.
In the SIDM models, the total velocity distributions are noticeably shifted to lower speeds compared to CDM.
This is related to the mass redistribution within the inner regions of the halo, especially in the  core region (see Fig. \ref{fig:veldisp}).
The largest suppression occurs for the inelastic SIDM model with \chih{}: compared to CDM, the mean speed decreases from 187.3~\kms{} to 168.5~\kms{}, while the median speed decreases from 179.4~\kms{} to 159.4~\kms{} (see Table \ref{tab:solarvel}).
Smaller changes are observed for elastic SIDM as well as for the cases with the smaller primordial excited fraction.

At higher velocities ($v \gtrsim$ 200~\kms{}), the SIDM models suppress the high-speed tail and result in a more steeply falling distribution, similar to that observed in some hydrodynamic simulations with baryonic physics \citep[e.g.][]{Pillepich2014,Butsky16v462}.
While the elastic SIDM halo has a velocity distribution close to Maxwellian at the solar radius, the inelastic halo distribution deviates more significantly from Maxwellian. 
This results from the lower number of scattering events in the inelastic SIDM model compared to the elastic model: within the inner 10~kpc, there are twice (1.2 times) as many scattering events in the elastic model relative to the inelastic model with primordial excitation fraction \chih{} (50\%) \citepalias[see the lower panels of Fig. 7 of][]{Vogelsberger19v484}.

The distributions of individual particle states generally follow that of the overall distribution, with the exception of the inelastic model with primordial excited  fraction \chih{} (left panel).
In this particular case, the model predicts a substantial fraction of ground state particles (red dotted line) with high speeds up to $v\sim 500$~\kms{} at the solar radius.
This result is coherent with the radial phase-space histogram (Fig. \ref{fig:phasespacehist}), where the presence of these high-speed particles previously been noted.
For the lower primordial excited fraction \chif{}, the phase-space histogram suggests the presence of similar high-speed ground state particles, 
but these are suppressed in the velocity distribution since such particles only constitute a tiny fraction of the ground state population (see Fig.~\ref{fig:denslope}).

%~~~~~~~~~~~~~~~~~~~~~~~~~~~~~~~~~~~~~~~~~~~~~~~~~~~~~~~~~~~~~~~~~~~~~~~~~~~

\subsection{Halo shape}

\begin{figure}
    \centering
	\includegraphics[width=0.47\textwidth]{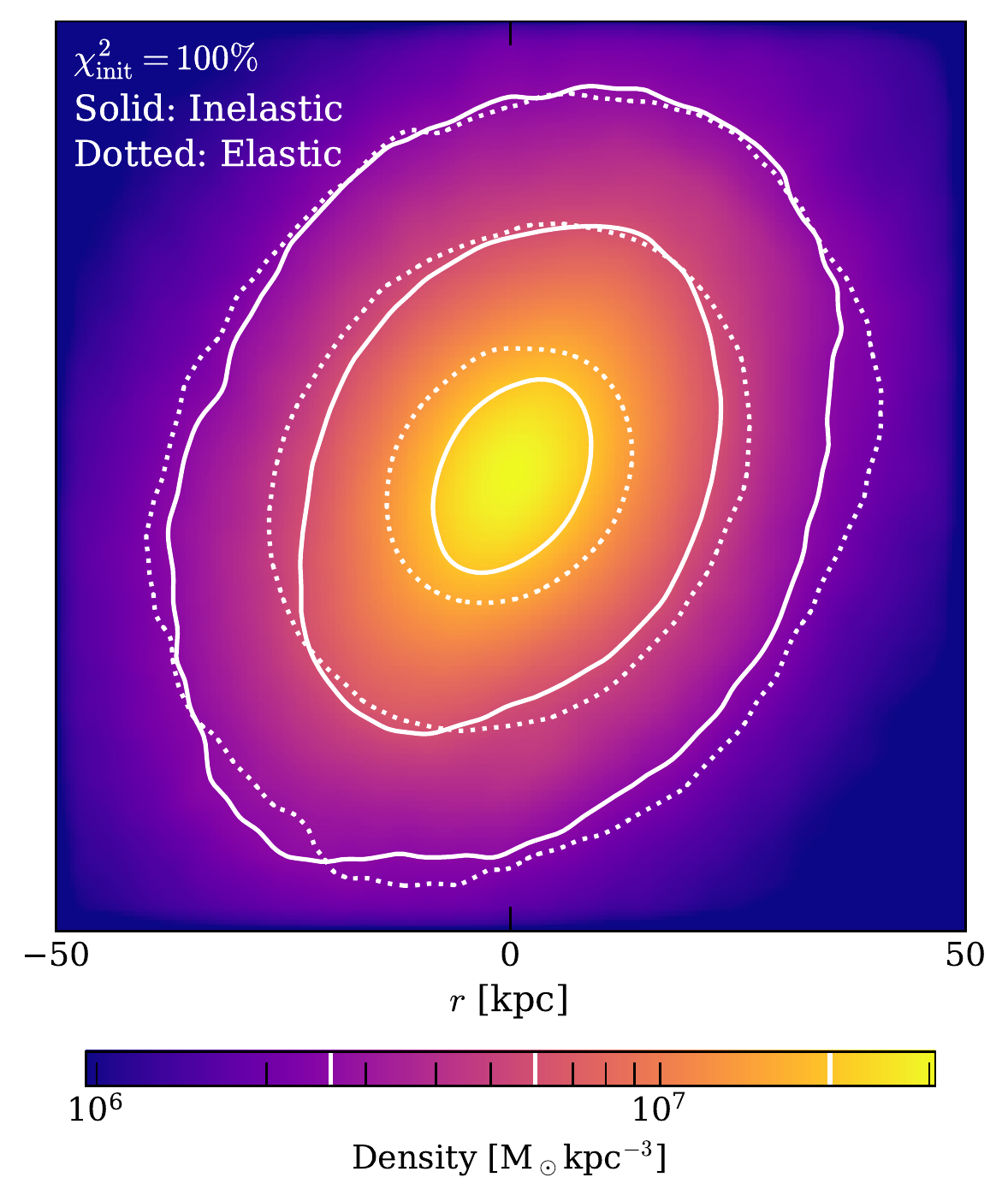}
    \caption{{\bf Dark matter density of the MW-size halo simulated with inelastic SIDM} for a primordial excitation fraction of \chih{}, within a 20~kpc-thick slice.
    To better elucidate the shape of the DM density, we show in solid lines three isodensity contours, which correspond to three halo-centric distances.
    For comparison, the corresponding isodensity contours for the elastic SIDM halo with the same excitation fraction are also shown as dotted lines.
    The isodensity contours show that the inelastic halo is less spherical than the elastic counterpart at these radii.
    }
    \label{fig:shapevis}
\end{figure}

\begin{figure*}
    \centering
	\includegraphics[width=\textwidth]{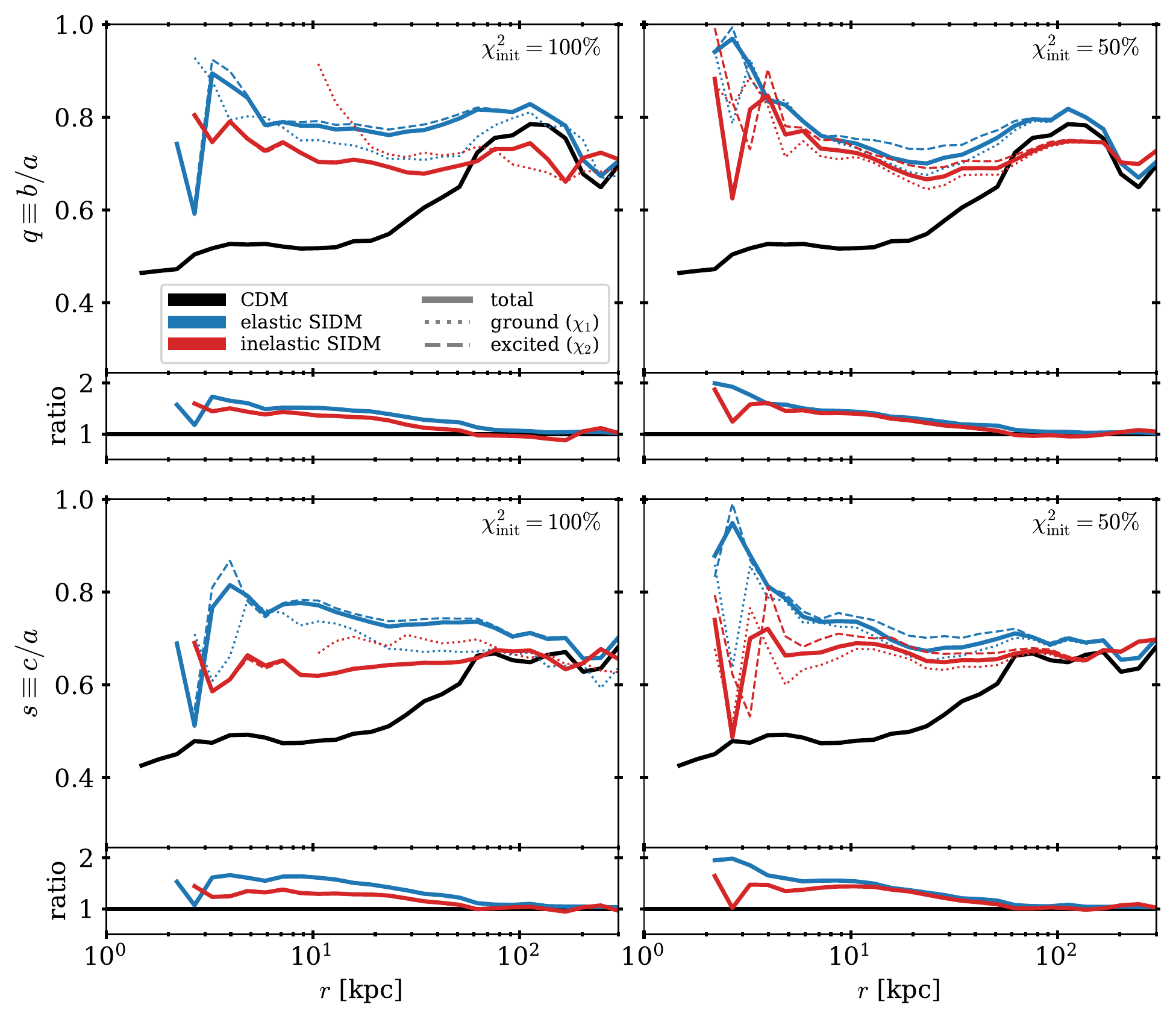}
    \caption{{\bf Halo shape profiles for the Milky Way-size halo}, showing $q \equiv b/a$ (top row) and $s \equiv c/a$ (bottom) as a function of radius.
    The bottom plots attached to each panel show the ratio of the SIDM profiles relative to CDM.
    All SIDM models lead to more spherical haloes (larger $q$ and $s$) than CDM for a significant fraction of the halo. 
    This increase in sphericity is strongest in the inner regions and decreases towards the virial radius.
    In the elastic models, the halo with \chih{} (left panels) is more spherical than its counterpart with \chif{} (right panels).
    The inelastic SIDM models lead to a less spherical halo compared to their elastic SIDM counterparts, especially for \chih{}.
    }
    \label{fig:shape}
\end{figure*}

Hierarchical structure formation theory and CDM simulations predict that dark matter haloes are triaxial, due to the anisotropic accretion of matter during halo growth \citep{Dubinski91v378,Warren92v399,Bullock2002,Jing02v574,Bailin05v627,Allgood06v367,Maccio08v391}.
Since self-interactions isotropise the velocities of the dark matter particles (see Section~\ref{sec:veldisp}), we expect SIDM to have an impact on halo shapes as well. 
The non-spherical nature of the dark matter halo is illustrated in Fig.~\ref{fig:shapevis}, which visualizes the dark matter density of the inelastic SIDM halo with primordial excitation fraction \chih{}, for $r<50$~kpc.
We also include three isodensity contours for the same model (solid lines), as well as for the elastic SIDM counterpart (dotted lines) to help elucidate the halo shape.

To quantify and compare the departure from spherical symmetry, we calculate the halo shape profiles using an iterative algorithm \citep{Allgood06v367,Zemp11v197}, which we briefly describe as follows. 
We assume that the isodensity surfaces can be described by ellipsoidal shells, and are interested in determining the halo shape at a particular radius through the axis ratios $b/a$ and $c/a$, where $a > b > c$ are the lengths of the principal axes of the shell.
For a set of particles with mass $m_k$ within a particular ellipsoidal shell, we define the components of the shape tensor as
\begin{equation}
    S_{ij} = \frac{\sum_ {k=1}^{N_p}  m_k x_k^{(i)} x_k^{(j)}}{ \sum_{k=1}^{N_p} m_k},
    \label{eqn:shapetensor}
\end{equation}
where $N_p$ is the number of particles within the shell and $x^{(i)}_k$ refers to the $i$th component of the $k$th particle.
Note that the shape tensor is directly proportional to the second moment of the mass distribution.
For each iteration, we compute the eigenvalues and eigenvectors of the shape tensor $S_{ij}$. The eigenvectors denote the orientation of the principal frame, while the eigenvalues ($\lambda_1 > \lambda_2 > \lambda_3$) give the axis ratios $q \equiv b/a = \sqrt{\lambda_2/\lambda_1}$ and $s \equiv c/a = \sqrt{\lambda_3/\lambda_1}$.
In the next iteration, we deform the ellipsoidal shell using the new values of $q$ and $s$ while keeping the length of the major axis constant. 
This means that only particles with elliptical radius $r_{\rm ell} = \sqrt{x^2 + y^2/q^2 + z^2/s^2}$ that fall within the bin width are used to calculate the new shape tensor. 
For a given elliptical radius $r_{\rm ell} = R$, we define the shell to be $0.85R < r_{\rm ell} < 1.15R$, thus shells are allowed to overlap and particles can belong to multiple shells simultaneously.
This procedure has been shown to produce reliable estimates of the halo shape \citep[e.g.][]{Zemp11v197, Brinckmann2018, Chua19v484}, although it differs slightly from other methods which keep the enclosed volume of the shell constant, or calculate shapes using the entire enclosed mass within the ellipsoidal volume.

We calculate the axis ratio profiles in 30 ellipsoidal shells spaced logarithmically between 1~kpc $< r <$ 300~kpc. To begin the algorithm, we start with spherical shells i.e. $q = s = 1$. The iterations are performed until convergence is obtained, when successive values of $q$ and $s$ differ by less than one per cent.  
%From the axis ratios $q$ and $s$, we calculate the triaxiality parameter $T \equiv (1-q^2)/(1-s^2)$.
%Haloes are considered to be completely prolate when $T = 1$, completely oblate when $T = 0$ and triaxial for $0.33 < T < 0.66$.

Previous convergence studies have shown that convergence in the DM halo shapes is more demanding that for the DM density profiles. 
Within a halo, $\kappa(r)$, the ratio of the two-body relaxation time-scale to the circular orbit time-scale at the virial radius, can be expressed as:
\begin{equation}
    \kappa(r) \equiv \frac{\sqrt{200}}{8} \frac{N(r)}{\ln N(r)} 
    \left[ \frac{\bar \rho(r) }{\rho_{\rm crit}} \right]^{-1/2},
\end{equation}
where $N(r)$ is the number of particles enclosed within radius $r$ and $\bar \rho(r)$ is the mean density within radius $r$ \citep{Power2003v338}.
The analysis of  MW-size haloes from the Aquarius simulation \citep{Aquarius,Vera-Ciro2011} as well as haloes from the Illustris project \citep{Chua19v484} have found that the convergence radius $r_{\rm conv}$, defined by $\kappa(r_{\rm conv}) = 7$, gives a good indication of the minimum radius where the shape profiles remain reliable in $N$-body simulations.
This choice of $r_{\rm conv}$ is larger than the Power radius $r_p$, defined by $\kappa(r_p) = 1$, which is traditionally applied to halo circular velocity profiles.
%In practice this estimate of $r_{\rm conv}$ serves as a conservative choice for the minimum radius for the SIDM models, since the SIDM cores are more easily resolved than the density cusps of CDM haloes.
In addition, we also consider only shells containing at least 1000 particles, which has been found to be approximately the minimum number of particles required for the iterative shape algorithm to be reliable \citep[e.g.][]{Tenneti14v441}.

We show the obtained shape profiles $q(r)$ and $s(r)$ from our simulations in Fig. \ref{fig:shape}.
For each profile, only the converged region ($r>r_{\rm conv}$) is shown.
From $q$ and $s$ (top and middle rows respectively), we find a clear separation between the SIDM cases and CDM for $r \lesssim 50$~kpc: all the SIDM cases result in dark matter haloes which are more spherical (larger $q$ and $s$) than the CDM model.
For the primordial excited fraction \chif{}, the elastic and inelastic SIDM models both have $q \approx 0.75, s \approx 0.7$ at a radius of $r=10$~kpc, much higher than the values $q=0.55, s=0.45$ for CDM.
Self-interactions tend to isotropise particle orbits, making haloes more spherical. 
This effect is strongest near the centre and decreases towards the viral radius where the SIDM shapes are very similar to that of CDM.

Interestingly, the inelastic SIDM model results in haloes that are less spherical than their elastic counterparts. 
For \chih{}, Fig.~\ref{fig:shape} shows that $s = 0.75$ for the elastic halo, compared to $s = 0.6$ for the inelastic halo at $r = 10$~kpc.
Visually, this result can also be observed by comparing the isodensity contours of the two models shown in Fig.~\ref{fig:shapevis}. 
This separation in the halo shape profiles persists up to 200~kpc, and provides an additional diagnostic which can potentially be used to distinguish between the elastic and inelastic SIDM models.
The larger sphericity in the elastic SIDM case can be explained by its larger number of scattering events compared to the inelastic halo, (see Section \ref{sec:solarvel} and Fig 7. of \citetalias{Vogelsberger19v484}).

For elastic SIDM, DM halo shapes have been previously studied by \cite{Peter13v430} and \cite{Brinckmann2018}.
Our elastic results are qualitatively similar to the median results of the galaxy-size haloes studied by \cite{Peter13v430}, when compared to the largest self-interaction cross section they considered ($\sigma/m = 1~\text{cm}^2\,\text{g}^{-1}$).

For the MW, estimates of the shape of the inner halo ($r \lesssim 30$~kpc) can be inferred observationally.
For example, stellar kinematics together with equilibrium modelling with the Jeans equations has suggested that $s=0.47\pm0.14$ \citep{Loebman2012}.
Tidal stellar streams are also widely used to model the MW halo shape, providing estimates of:
$s$~>~0.7 \citep{Ibata01v551}, $s$~=~0.72 \citep{Law10v714}, $s$~=~0.8 \citep{Vera-Ciro13v773}, and $s=$~1.05~$\pm$~0.14 \citep{Bovy16v833}.
All the SIDM haloes, being more spherical, are in better agreement with the observational values compared to CDM.
While the observations appear to favour the more spherical elastic model, it is important to note that an exact comparison with the MW halo shape must necessarily incorporate the dynamical impact of the assembly of the luminous galaxy, which is not taken into account in our simulations.

%~~~~~~~~~~~~~~~~~~~~~~~~~~~~~~~~~~~~~~~~~~~~~~~~~~~~~~~~~~~~~~~~~~~
\section{Assembly History}

\begin{figure*}
	\includegraphics[width=0.95\textwidth]{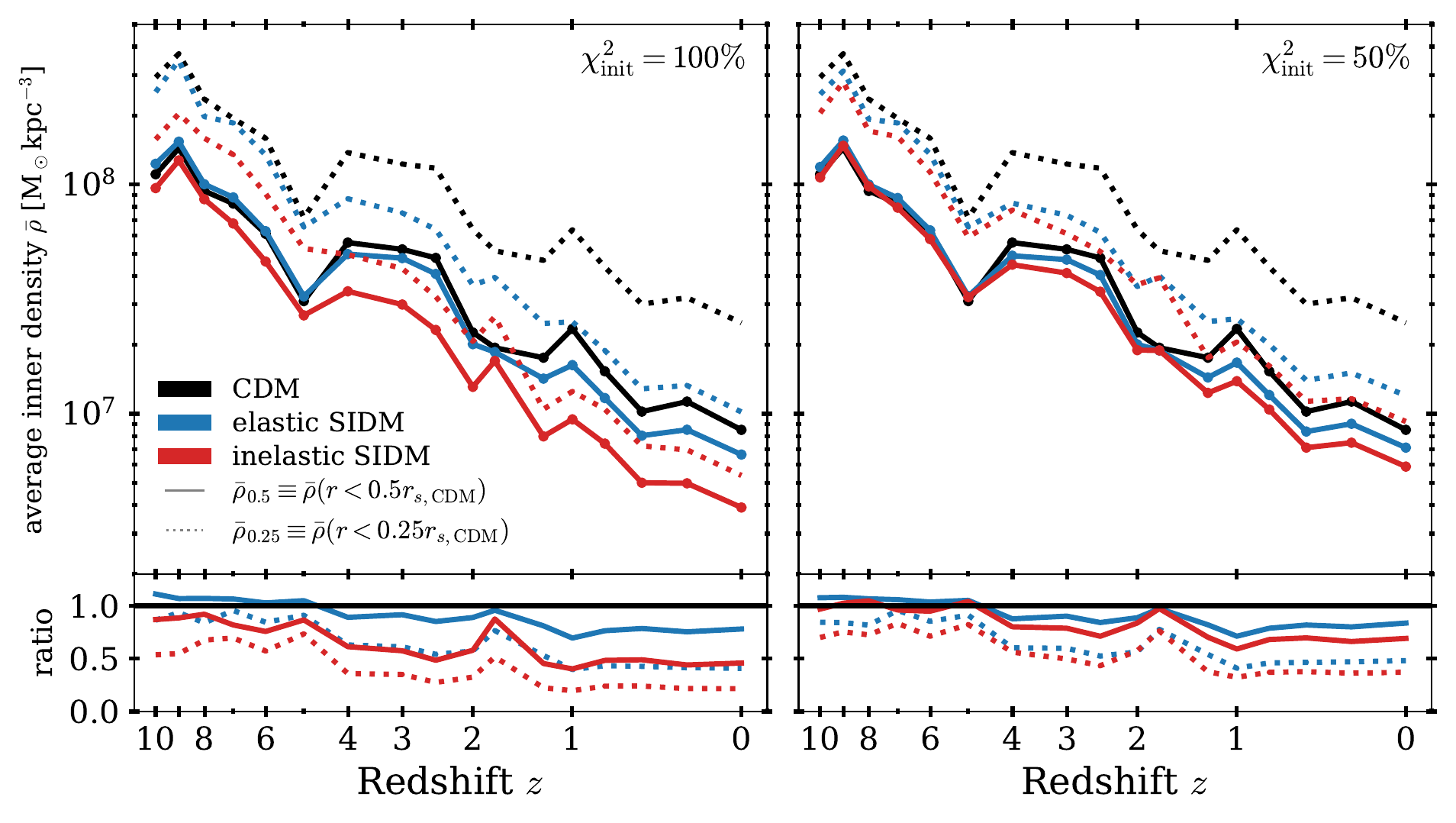}
    \caption{\textbf{Average inner density (in physical units) as a function of redshift for the MW-like halo}, within $0.25r_{s,\rm CDM}$ and $0.5r_{s,\rm CDM}$.
    At each redshift, the scale radius of the CDM halo $(r_{s,\rm CDM})$ is used across the models for consistency.
    The bottom plots attached to each panel show the ratio of the SIDM quantities relative to CDM.
    For a fixed primordial excitation fraction, the inner densities are decreased in a shorter timescale in the inelastic SIDM model compared to the elastic model, thus the inelastic halo has the lowest density at each redshift. 
    At $z=10$,  $\bar \rho_{0.5}$ is suppressed relative to CDM only in the inelastic model with \chih{}, due to the larger enclosed mass within $0.5r_{s,\rm CDM}$, 
    }
    \label{fig:red_dense14}
\end{figure*}

\begin{figure*}
	\includegraphics[width=0.95\textwidth]{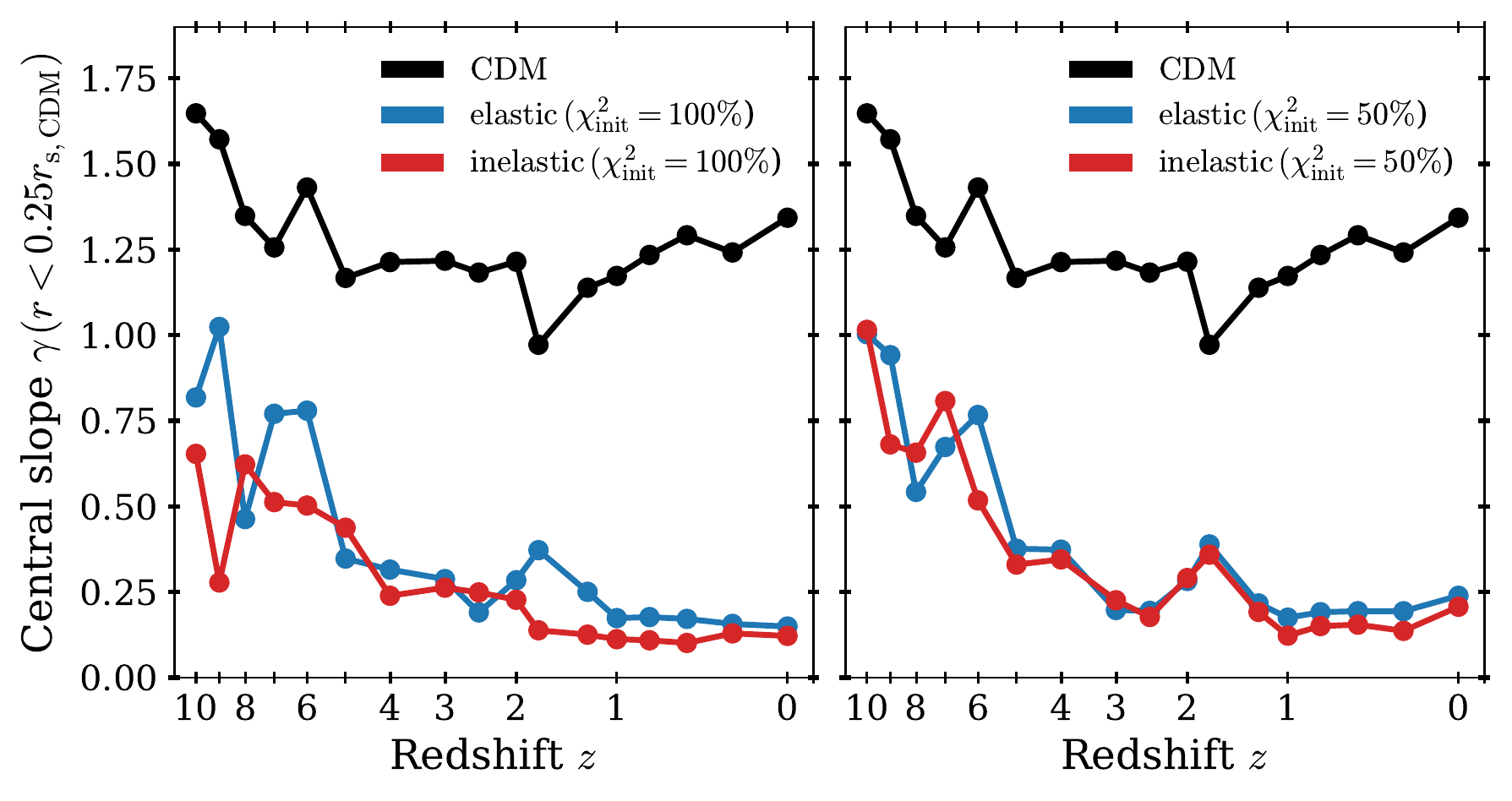}
    \caption{\textbf{Inner slope as a function of redshift for the MW-like halo}.
    At each snapshot, the inner slope $\gamma$ is obtained from fitting the density profile $\rho(r)$ within 1/4 of the NFW scale radius of the CDM halo counterpart to a power law $r^{-\gamma}$.
    For the CDM case, $\gamma$ is approximately constant across redshift, remaining at $\gamma \approx 1.25$.
    In all the  SIDM cases, self-interactions already result in a flattened density profile at redshift $z=10$ compared to the CDM counterpart.
    Between $z=10$ and $z=0$, the inner slope becomes less steep due to the increasing impact of self-interactions, with $\gamma$ decreasing roughly from 1 to 0.25 in all the SIDM cases.
    At low redshifts, the inelastic halo is slightly less steep than the elastic counterpart, due to the larger core formation efficiency in the former.
    }
    \label{fig:red_slope14}
\end{figure*}

% \begin{figure*}
% 	\includegraphics[width=0.95\textwidth]{Figures/slope_z_35.pdf}
%     \caption{\textbf{Slope parameter $b$ as a function of redshift for the MW-like halo}.
%     At each snapshot, the parameter $b$ is obtained from fitting $\Gamma \equiv -d\ln\rho/d\ln r$ to a power law $r^b$ within the half the NFW scale radius of the CDM halo counterpart.
%     In general, $b$ increases with time (decreasing redshift) between $z=10$ and $z=4$, remaining approximately constant thereafter.
%     At low redshifts (near $z=0$), the parameter $b$ is $b\approx 0$ in the CDM model,  which is consistent with an NFW profile. 
%     In the SIDM models however, the parameter $b$ is much larger relative to CDM at all the examined redshifts, indicating that both elastic and inelastic self-interactions have already impacted the dark matter distribution even at $z=10$.
%     }
%     \label{fig:red_slope14}
% \end{figure*}

So far, we have discussed the structural characteristics of inelastic SIDM haloes only at redshift $z=0$. 
It is also relevant to consider how these characteristics change as the haloes evolve from high redshift to the present epoch.
%Following \cite{Brinckmann2018}, we focus on the  region $r < 0.5r_{s,\rm CDM}$, where $r_{s,\rm CDM}$ is the scale radius of the CDM halo.

For a CDM halo, the density profile is well-described by an NFW profile \citep{Navarro1996}:
\begin{equation}
 \frac{\rho_{\rm NFW}(r)}{\rho_\text{crit}}= \frac{\delta_c}{(r/r_s) \left(1+r/r_s\right)^2},
 \label{eqn:nfw}
\end{equation}
where $\delta_c$ is a characteristic density contrast and the scale radius $r_s$ is the radius where the logarithmic density slope is $d\ln\rho/d\ln r = -2$.
Due to the presence of the constant density core in the SIDM haloes, the SIDM density profiles are not well-described by the NFW profile  (see Section~\ref{sec:density} and also \citealt{Todoroki19v483b}).
Thus, we only calculate $\delta_c$ and the scale radius $r_{s,\rm CDM}$ for the CDM halo using a two-parameter least-squares fit.
At redshifts $z=0$ and $z=10$, the comoving scale radii for the CDM halo are $r_{s,\rm CDM} = 24.0$~kpc and 14.8~kpc, respectively.
%Plummer equivalent softening lengths in comoving units are 72.4pc and 724pc, respectively.

We first examine the evolution of the average inner density within $0.25 r_{s,\rm CDM}$ and $0.5 r_{s,\rm CDM}$.
This results in the two measures: (i) $\bar \rho_{0.25} \equiv \bar  \rho(r<0.25r_{s,\rm CDM})$, 
and (ii) $\bar \rho_{0.5} \equiv \bar \rho(r<0.5r_{s,\rm CDM})$.
For each of the five models, we trace the main halo from the final snapshot at $z = 0$ back to $z = 10$, picking in each snapshot its progenitor based on mass and distance.
Subsequently, $\bar \rho_{0.25}$ and $\bar \rho_{0.5}$ are calculated for each redshift using the scale radius of the CDM counterpart. 
The standardisation of the radius (i.e. using $r_{s,\rm CDM}$)  at each redshift ensures that the inner densities calculated  can be compared consistently across the models.

Fig.~\ref{fig:red_dense14} shows the evolution of $\bar \rho_{0.25}$ and $\bar \rho_{0.5}$ in physical units for CDM and the SIDM models, with the bottom attached plots showing the ratio of the inner densities to that of the CDM model.
For $\bar \rho_{0.25}$, the effect of dark matter self-interactions manifests with time (decreasing redshift), and increasingly suppresses the inner density relative to CDM with time.
Relative to the CDM halo, $\bar \rho_{0.25}$ is suppressed by 15--45 per cent at $z=10$ across the SIDM models.
By $z=0$, this number has increased to 50--80 per cent.
At fixed primordial excitation fraction \chii{}, the inelastic halo has the lowest inner density at each redshift.
Thus, we conclude that energy injection from inelastic down-scattering reduces the inner density in a shorter timescale compared to the elastic scale.

The effects of self-interactions are weaker on $\bar \rho_{0.5}$ compared to $\bar \rho_{0.25}$, due to the larger enclosed mass within $0.5 r_{s,\rm CDM}$.
At $z=0$, $\bar \rho_{0.5}$ is only suppressed by 15--55 per cent relative to CDM.
At high redshifts, due to the smaller cumulative number of scattering events, only an inelastic model with high primordial excitation fraction injects sufficient energy through inelastic interactions to evacuate dark matter from within $0.5r_{s,\rm CDM}$.
As a result, only the inelastic SIDM model with \chih{} (left panel, red curve) shows a substantial decrease in  $\bar \rho_{0.5}$  relative to CDM (-14 per cent).

Finally, we examine the inner slope of the logarithmic density profile, which provides more information about the growth of the constant density core. 
Here, we measure the inner slope $\gamma$  by fitting the density profile in the region $r<0.25 r_{s, \rm CDM}$ to a power law $\rho(r) \propto r^{-\gamma}$.
We emphasise again that at each redshift, the scale radius for the CDM halo is used for consistency across the models.
The resulting evolution of the inner slope $\gamma$ with redshift is shown in Fig.~\ref{fig:red_slope14}.
In the CDM model, we find the inner slope is approximately constant across redshift, with a value of $\gamma \approx 1.25$.
This value is consistent with the NFW profile\footnote{
For the NFW profile, the local slope of the logarithmic density profile  is $\Gamma(r) \equiv  -d\ln\rho /d\ln r = 1+ 2r/(r+r_s)$.}, 
for which the local slope is $\Gamma(r=0) = 1$ at the centre and  $\Gamma(r_s) = 2$ at the scale radius.

In the SIDM models, both elastic and inelastic self-interactions already flatten the density profiles at redshift $z=10$ compared to the CDM model.
Between $z=10$ and $z=0$, $\gamma$ decreases from $\gamma \approx 1$ to $\gamma \approx 0.25$ in the SIDM cases.
As the halo grows, self-interactions cause the inner core to grow, which leads to a shallower inner slope.
At low redshifts, the inner slopes in the inelastic cases are slightly smaller (i.e. the density profiles are flatter) than the elastic cases, due to more efficient core formation associated with inelastic SIDM.

Comparing these inner slopes with the central densities from Fig.~\ref{fig:red_dense14}, we note that at high redshifts, elastic self-interactions alone have only a small effect on the average inner densities $\bar \rho_{0.5}$ and $\bar \rho_{0.25}$ relative to CDM.
However, the effect of  elastic and inelastic self-interactions on the inner slope is already considerable  even at $z=10$.

\section{Conclusions}

We have examined the results of  high resolution simulations of a Milky Way-size dark matter halo composed of inelastic self-interacting dark matter.
Our inelastic SIDM model consists of nearly degenerate two-state dark matter particles with an energy level splitting of $\delta=10$~keV, which we implemented and simulated using {\sc Arepo}.
During the exothermic down-scattering process, the model results in velocity kicks of approximately 424~\kms{}.
To understand the effects of such energy injection, we have examined and compared inelastic SIDM simulations of a MW-size halo against an elastic SIDM model and the conventional CDM model.
The elastic SIDM model simulated in this work suppresses the energy change during down-scattering and up-scattering but is otherwise identical to inelastic SIDM.
In addition, we have also distinguished between the configurations where the simulation begins with all DM particles in the excited state (\chih{}) and where only half begin in the excited state (\chif{}).
Using these five cases, we examined the effects of elastic and inelastic SIDM on the internal structure and assembly of dark matter haloes.
In the following, we summarise our main findings, concentrating on the case with primordial excited fraction \chih{}:

\begin{enumerate}

\item  Energy injection resulting from inelastic self-interactions reduces the central density of the inelastic SIDM halo in the inner regions of the halo and results in a larger core compared to the elastic counterpart  (Fig.~\ref{fig:denslope}).
In the \chih{} configuration, inelastic SIDM reduces the density by a factor of approximately 20 relative to CDM, compared to a factor of 10 for elastic SIDM, at a radius $r = 1$~kpc.
Although the density profiles in the inner 3~kpc are flat for both elastic and inelastic SIDM, the logarithmic slope in the elastic models approaches that of CDM more rapidly with increasing radius.

\item At intermediate radii, the inelastic SIDM radial density profiles are suppressed relative to CDM, because the velocity kicks from inelastic down-scattering unbind and eject particles from the MW halo.
In contrast, the density profiles of the elastic SIDM models are slightly raised compared to CDM since particles are solely transferred from the inner regions to the intermediate regions of the halo.
The density of the ground state particles in the inelastic model is around two orders of magnitude lower in the core region compared to the elastic model, reflecting the characteristic particle ejection unique to inelastic SIDM.

\item We found that self-interactions flatten the velocity dispersion in the inner regions (Fig.~\ref{fig:veldisp}, top panels), causing the halo to become isothermal, in contrast to the ``temperature inversion" observed in the CDM case.
Inelastic self-interactions leads to lower velocity dispersions compared to elastic interactions, due to the expulsion of high-speed particles from the inelastic halo.

\item Inelastic scattering results in an inelastic halo which is more isotropic ($\beta \approx 0$) than the CDM and elastic SIDM counterparts (Fig.~\ref{fig:veldisp}, lower panels).
Although elastic SIDM results in a similar isotropisation in the inner and intermediate regions of the halo, it does not modify the velocity anisotropy substantially in the outer regions.
This implies the energy released in the exothermic interactions could be important in modifying particle orbits near the halo outskirts.

\item Self-interactions flatten the pseudo phase-phase density (Fig.~\ref{fig:phasespace}) in the inner regions, resulting in a profile substantially different from the universal power law followed by CDM haloes.
Further analysis of the radial phase-space distributions (Fig.~\ref{fig:phasespacehist}) reveals a substantial population of ground state particles with radial velocities of up to 500~\kms{} which is uniquely present in the inelastic models.
These particles are unbound from the central subhalo and reflect the velocity kicks associated with inelastic down-scattering.

\item The local velocity distribution $f(v)$ of DM particles at the solar circle (Fig.~\ref{fig:solarvel}) shows that the inelastic SIDM models predict the presence of high-speed ground state particles which have received velocity kicks.
The decreased core density in the inelastic models results in a total $f(v)$ that differs from both CDM and elastic SIDM.
In the SIDM haloes, $f(v)$ is shifted to lower velocities, together a suppression of the high-speed tail ($v \gtrsim 200$~\kms{}).
For inelastic SIDM, we also found that the ground state particles show a distinct population of high-speed particles ($v \gtrsim 400$~\kms{}). 
These particles are not present in the velocity distribution of the excited state particles or in the elastic SIDM counterpart.
The unique presence of high-speed particles in velocity distribution $f(v)$ of  the inelastic SIDM halo is a potential signature for direct detection experiments.

\item Dark matter self-interactions result in haloes that are more spherical than the CDM counterpart (Fig.~\ref{fig:shape}), and are thus in better agreement with observational estimates of the MW halo shape.
Interestingly, we found that our simulated halo was more spherical with elastic SIDM than with inelastic SIDM.
For example, the minor-to-major axis $s$ ratio at $r = 10$~kpc is $s = 0.6$ for the inelastic halo, compared  to $s = 0.75$ for the elastic halo and $s = 0.45$ for the CDM halo.
The larger number of scattering events in the elastic halo drives the shape to more more spherical compared to the inelastic counterpart.

\item Tracing the halo assembly history, we found that inelastic self-interactions reduce the inner density of the MW halo in a shorter timescale relative to the elastic scale (Fig.~\ref{fig:red_dense14}), and also result in a shallower inner slope at low redshifts (Fig.~\ref{fig:red_slope14}).
As such, inelastic self-interactions affect significantly the innermost region of the halo ($r\lesssim0.25r_s$) at high redshifts ($z\gtrsim5$).
We found that both elastic and inelastic self-interactions already result in a substantial flattening of the central density profile, even at a redshift as high as $z=10$.
\end{enumerate}

For simulations carried out with the lower primordial excitation fraction (\chif{}), the results are in general similar to that observed for \chih{}.
With a lower \chii{}, the differences relative to CDM were more subdued because:
(i) less energy is injected into the halo when less particles start out in the excited state,
and (ii) the down-scattered ground state particles form only a small fraction of the total ground state particles in the halo.
While energy injection from inelastic self-interactions drives many of the differences  between the inelastic and elastic haloes, the number of scattering events appears to be the primary factor for certain situations.
For example, the latter factor is responsible for the local velocity distribution $f(v)$ of the elastic SIDM haloes being closer to Maxwellian than their inelastic counterparts, and also for the elastic haloes being more spherical.

In conclusion, we have found that inelastic self-interactions can significantly impact dark matter haloes, causing the structure of inelastic SIDM haloes to exhibit distinct differences with respect to their elastic SIDM and CDM counterparts.
By performing simulations with only dark matter particles, we have focused on the effects of elastic and inelastic scattering through comparisons to the CDM model.
Although we explored two primordial excitation fractions, we remark that our results presented in this work correspond to a single choice for the parameters $\delta$, $\alpha$, $m_{\chi}$ and $m_{\phi}$ in the \cite{Schutz2015} model.
In future work, we plan to expand the study by exploring different choices in the parameter space.
At the same time, it is well known that hydrodynamics simulations with CDM and baryons have predicted that baryonic physics associated with galaxy assembly e.g. gas cooling, stellar feedback, black hole feedback etc. has a relevant impact on the structure of dark matter haloes.
Future simulations incorporating both inelastic SIDM and galaxy formation physics could prove to be helpful in understanding their combined effects on both the dark matter halo as well as the luminous galaxy.

%~~~~~~~~~~~~~~~~~~~~~~~~~~~~~~~~~~~~~~~~~~~~~~~~~~~~~~~~~~~~~~~~~~~~~~~~~~~~~~~~~

\section*{Acknowledgements}

We thank David Barnes for his constructive comments on the paper.
The simulations were performed on the joint MIT-Harvard computing cluster supported by MKI and FAS.
JZ acknowledges support by a Grant of Excellence from the Icelandic Research Fund (grant number 173929).

\section*{Data Availability}

The data underlying this article will be shared on reasonable request to the corresponding author.

%%%%%%%%%%%%%%%%%%%% REFERENCES %%%%%%%%%%%%%%%%%%
\FloatBarrier

% The best way to enter references is to use BibTeX:
\bibliographystyle{mnras}
\bibliography{references} % if your bibtex file is called example.bib

%%%%%%%%%%%%%%%%%%%%%%%%%%%%%%%%%%%%%%%%%%%%%%%%%%

%%%%%%%%%%%%%%%%% APPENDICES %%%%%%%%%%%%%%%%%%%%%

%%%%%%%%%%%%%%%%%%%%%%%%%%%%%%%%%%%%%%%%%%%%%%%%%%

% Don't change these lines
\bsp	% typesetting comment
\label{lastpage}
\end{document}